\documentclass{rmaa}
\suppressfulladdresses
\usepackage{paralist}
\usepackage{lscape}
\SetYear{2006}

\title{The properties of optical FeII emission lines of AGN with
double-peaked broad emission lines}

\author{Xue-Guang, Zhang\altaffilmark{1,2},
       Deborah, Dultzin-Hacyan\altaffilmark{1},
        Ting-Gui, Wang\altaffilmark{2}}
\altaffiltext{1}{Instituto de Astronomia, Universidad Nacional Autonoma de
                 Mexico, Apdo Postal 70-264, Ciudad Universitaria, 
                 Mexico D. F. 04510, Mexico \\
                 xguang@astroscu.unam.mx, deborah@astroscu.unam.mx}
\altaffiltext{2}{Center for Astrophysics, Department of astronomy and Applied
                 Physics, University of Science and Technology of China, Hefei,
                 Anhui, P.R.China\\
                 twang@ustc.edu.cn}

\shorttitle{The properties of optical FeII emission lines of dbp emitters}
\shortauthor{Zhang, Dultzin-Hacyan \& Wang}
\listofauthors{Zhang, X-G, Dultzin-Hacyan, D. \& Wang, T-G}
\indexauthor{Zhang, X-G}
\indexauthor{Dultzin-Hacyan, D.}
\indexauthor{Wang, T-G}

 \resumen{
    Estudiamos las propiedades de una muestra de 27 Nucleos Activos de
Galaxias (NAGs) con perfil de doble pico en las lineas anchas de
emision de baja ionizacion, tomados del SDSS (DR4) con un valor
medio de $\sigma_{H\alpha_{B}}\sim3002\pm139{\rm km\cdot s^{-1}}$. Nuestro
primer resultado es que todo el espectro de emision en el rango de
4100$\AA$ a 5800$\AA$ se puede ajustar muy bien mediante un modelo
de emision en un disco de acrecion eliptico, suponiendo que las
lineas de H$\beta$, FeII, H$\gamma$ y HeII$\lambda4686\AA$ tienen el
mismo perfil de doble pico que las lineas de H$\alpha$. Un analisis
similar se ha hecho en un articulo previo para el objeto SDSS
J2125-0813. Los mejores ajustes indican que la emision optica de las
lineas de FeII en los emisores dbp (emisores de lineas de doble
pico, por sus siglas en ingles) se originan en la misma region del
disco que las lineas de Balmer. Encontramos que algunas
correlaciones  conocidas para los NAGs "normales" se cumplen tambien
para los dbp. Sin embargo, este resultado debe tomarse con cautela
porque el numero de objetos es pequeño y porque tenemos un sesgo de
seleccion hacia objetos con emision de FeII intensa. Mostramos que
para los dbp, la masa del agujero negro central parece tener mayor
influencia en las propiedades de la emision de FeII que la taza de
acrecion. Tambien encontramos que todos los dbp de nustra muestra
son cuasares radio callados, con excepcion de ocho objetos con
$R_r> 1$ de acuerdo a la definicion de Ivezi$\rm{\acute{c}}$ et al.
(2002) y 6 objetos no descubiertos por FIRST.}

 \abstract{
    We study the FeII properties of double-peaked broad low-ionization
emission line AGN (dbp emitters) using a sample of 27 dbp emitters
from SDSS (DR4), with mean value
$\sigma_{H\alpha_{B}}\sim3002\pm139{\rm km\cdot s^{-1}}$. Our first
result is that the line spectra in the wavelength range from
4100$\AA$ to 5800$\AA$ can be best fitted by an elliptical accretion
disk model, assuming the same double-peaked line profiles for
H$\beta$, FeII, H$\gamma$ and HeII$\lambda4686\AA$ as that of
double-peaked broad H$\alpha$ for all the 27 dbp emitters, except
the object SDSS J2125-0813 which we have discussed in a previous
paper. The best fitted results indicate that the optical FeII
emission lines of dbp emitters originate from the same region in the
accretion disk where the double-peaked Balmer emission lines
originate. Some correlations between FeII emission lines and the
other broad emission lines for normal AGN can be confirmed for dbp
emitters. However, these results should be taken with caution due to
the small number of objects and the bias in selecting strong FeII
emitters. We show that for dbp emitters, BH masses seems to have
more influence on FeII properties than dimensionless accretion rate.
We also find that the dbp emitters in the sample are all radio quiet
quasars except one dbp emitter with $R_r> 1$ according to the
definition by Ivezi$\rm{\acute{c}}$ et al. (2002) and 6 objects
undiscovered by FIRST. }

\addkeyword{Galaxies:Active}
\addkeyword{Quasars:Emission lines}
\addkeyword{accretion disk}

\begin{document}
\maketitle

\section{Introduction}

   The optical FeII emission lines are one of the most important properties
of some Active Galactic Nuclei (AGN). Because of the large amounts
of FeII emission lines (half-filled 3d-shell leads to thousands of
transitions for FeII emission lines) (Moore \& Merill 1968; Netzer
1988; Wills et al. 1980; Wills, Netzer \& Wills 1980), it is
difficult to clarify many properties of these lines, such as their
loci of emission, the mechanism of their excitation, the correlation
between FeII lines and other emission lines. The bright object I Zw
I (PG 0050+124) has very strong FeII emission lines which can be
accurately measured. And thus it has been used to build FeII
templates for the subtraction of these lines when reducing quasar
spectra in order to properly fit the continuum (Philips 1978; Oke \&
Lauer 1979; Boroson \& Green 1992; Laor et al. 1997; Marziani et al.
2003a).

{  A detailed study of line-related correlations between FeII and
other properties of AGN can be found in Sulentic, Marziani \&
Dultzin-Hacyan (2000): there is an outstanding strong
anticorrelation between EW(FeII4570)/EW(H$\beta$) and FWHM($H\beta$)
in the context of the so called Eigenvector 1 (E1) parameter space.
These authors have defined a population A,
FWHM(H$\beta_{broad}$)$<4000{\rm km\cdot s^{-1}}$, and population B,
FWHM(H$\beta_{broad}$)$>4000{\rm km\cdot s^{-1}}$, which is a
cleaner distinction for AGN with strong and weak FeII emission lines than
Radio loud objects vs. Radio quiet objects. A preliminary definition
of Eigenvector 1 can be found in (Boroson \& Green 1992). An updated
and more complete version of it includes the interpretation of the
observed trends in this parameter space in the context of their
physical drivers: BH masses and accretion rate (Sulentic et al.
2006).}

   There are two main kinds of models which
can reproduce the observed shape and equivalent width of FeII
emission lines: Photoionization models with microturbulence
(Korisra et al. 1997; Bottorff et al. 2000; Netzer 1985;
Baldwin et al. 2004) in the context of the so called LOC model
(Baldwin 1995) and collisionally excited models (Grandi 1981,
1982; Kwan et al. 1995; Dumont et al. 1998; Baldwin et al. 2004). A
simple photoionization scheme is not adequate to reproduce the
observed FeII  lines (Bergeron \& Kunth 1980;
Collin-Souffrin et al. 1986; Joly 1987; Collin-Souffrin,
Hameury \& Joly 1988) because of the enhanced FeII emission due to
the needed high density ($>10^{11}{\rm cm^{-3}}$) of the emitting clouds
(Ferland \& Person 1989). Thus, the accretion disk and/or the region
near the center produced by shocks along the radio jet provide a natural
high density environment for FeII emission lines.  Sigut \& Pradhan
(2003) have examined this assumption for the typical physical conditions
of AGN. In their paper, the following excitation mechanisms
for FeII emission lines have been considered: continuum fluorescence
(Phillips 1978, 1979), collisional excitation (Joly 1991),
self-fluorescence among the FeII transitions and fluorescent excitation
by Ly$\alpha$ and Ly$\beta$ (Penston 1988; Sigut \& Pradhan 1998;
Verner et al. 1999).

    The most commonly accepted  outcome from the study of FeII emission
lines is that the line width of these lines is the same as that of
the other broad low-ionization emission lines such as H$\beta$,
except for several special objects such as HE 1249-0648 and HE
1258-0823 (FWHM($H\beta$)$>>$FWHM(FeII)) (Marziani et al. 2003a,
2003b). The best way to determine where do the FeII emission lines
originate, the accretion disk or the base of the jet, is to study
the properties of FeII lines from a special kind of AGN which emit
double-peaked broad low-ionization Balmer emission lines (dbp emitters). 
These double-peaked broad 
lines are believed to originate from the accretion disk near the
central Black Hole (BH) (Storchi-Bergmann, Nemmen, et al. 2003;
Storchi-Bergmann, Eracleous et al. 1997; Storchi-Bergmann,
Eracleous \& Halpern 1995; Eracleous et al. 1997;
Storchi-Bergmann, Baldwin et al. 1993; Chen et al. 1989a, 1989b,
1997; Halpern et al. 1996; Antonucci et al. 1996; Sulentic et al.
1990; Shapovalova et al. 2001; Gilbert et al. 1999; Hartnoll \&
Blackman 2002; Karas et al. 2001). Other models have been
considered to interpret the origination of doubloe-peaked broad lines, 
such as the Binary Black Hole model (Begelman et al.
1980; Gaskell 1983) and bipolar outflowing model (Zheng et al.
1990), but they have been proved unsuccessful to explain most
dbp emitters.

   In a previous paper we have reported a flat-spectrum radio quiet
quasar, SDSS J2125-0813, which has a broad double-peaked H$\beta$ (
FWHM($H\beta$)$\sim15000{\rm km\cdot s^{-1}}$), and strong FeII emission
lines at optical bands that have exactly the same line profile as that 
of the broad
double-peaked H$\beta$ line (Zhang, Dultzin-Hacyan \& Wang 2006a,
hereafter paper I). In this paper, we select a whole sample of AGN
with both double-peaked broad Balmer lines and strong optical FeII
emission lines to study the FeII properties of dbp emitters. There
are two famous samples of dbp emitters, one consists of 23 objects
which are nearly all LINERs from radio galaxies (Eracleous \&
Halpern 1994, 2003; Eracleous et al. 1995), and the other sample
is made up of 112 objects of which 12\% are classified as LINERs
(Strateva et al. 2003) selected from SDSS DR2 (York et al. 2000).
Recently, we have built a larger sample of dbp emitters from SDSS
DR4 (Zhang et al. 2006b), which includes more than three hundred
dbp emitters. Because of the convenience to get the spectra of SDSS,
we select dbp emitters with apparent FeII emission lines from the
two samples in SDSS (Strateva et al. 2003; Zhang et al. 2006b). In
section 2, we list the data of our sample. Section 3 presents the
results and then the discussions and conclusions follow in Section
4. The cosmological parameters $H_{0}=75~{\rm km~s}^{-1}{\rm
Mpc}^{-1}$, $\Omega_{\lambda}=0.7$ and $\Omega_{m}=0.3$ have been
adopted here.

\section{The Sample}

   There are 112 dbp emitters selected from SDSS DR2 by
Strateva et al. (2003) and more dbp emitters are being selected from
SDSS DR4 by Zhang et al. (2006b). Because we select dbp emitters
from the main catalogs of galaxies and of quasars classified by
pipeline of SDSS, with different criteria from those used by
Strateva et al. (2003), and because in this paper we simultaneously
select dbp emitters with apparent and strong FeII emission lines, even 
when we use DR4, our sample is smaller (some dbp emitters selected by
Strateva et al. (2003) are not included in our sample). 

  { The
detailed selction criteria will be described in a forthcoming paper
(Zhang et al. 2006b). We give here a description of the steps we
followed: First, we subtracted the stellar components from the
observed spectra by the PCA template method (Hao et al. 2005; Li et
al. 2005), if necessary. Then the emission lines, especially 
H$\alpha$+[NII]$\lambda6548,6583\AA$, 
are fitted by gaussian functions. At least two gaussian functions, one 
broad and one narrow, are applied for each permitted emission line, 
and one gaussian function for each forbidden emission line.
According to the best fitted results by Levenberg-Marquardt least-squares 
minimization technique, the AGN with broad emission lines are selected 
according to the following criteria: the line width ($\sigma$), and
flux of the broad component of H$\alpha$ should be at least 5 times
larger than the measured errors, and also the line width $\sigma$ of the broad 
component of H$\alpha$ should be larger than $800{\rm km\cdot s^{-1}}$.
The value of the summed squared residuals divided by degrees of freedom, 
$\chi^2_1$, can also be obtained. The value $\chi^2_1$ near to 1  
represents whether the parameters of the model are significant
for the fit to the emission lines.
Because of the complex line profiles of double-peaked emission line, 
one broad gaussian function is not the better choice to fit the broad 
component, the values of $\chi^2_1$ for objects with double-peaked 
broad H$\alpha$ should be larger than 1. Thus multi-gaussian functions: 
Four broad gaussian functions, are applied for broad emission lines again.
Then We obtain the other value of $\chi^2_2$ for each AGN with broad 
emission lines. For objects with double-peaked broad H$\alpha$, the 
value of $\chi^2_2$ should be much closer to 1. We then select 
our dbp emitter candidates according
to the following criterion: $\chi^2_1 - \chi^2
> 1$, which indicates the observed line profiles can not be properly
fitted by one broad gaussian function.
Finally, we check the dbp emitter candidates one by one by eye,
according to whether there is apparent double peaks. We reject the
candidates which have asymmetric line profiles of H$\alpha$ or have
multi-gaussian components with the same center wavelength (the
separation of peaks is less than $10\AA$). }

  According to the observed spectra of
SDSS, we can easily select the dbp emitters for which the FeII
emission lines are evident, especially in the wavelength range from
4100$\AA$ to 5800$\AA$. We find 27 dbp emitters with apparent FeII
emission lines including the object SDSS J2125-0813, { 12 dbp
emitters are selected from the sample of Strateva et al. (2003) and
15 new dbp emitters are selected from SDSS DR4.} We show the spectra
in Fig 1 (the spectra of SDSS J2125-0813 can be found in Paper I,
here we do not show it again). According to the properties of the
spectra, all the 27 dbp emitters are quasars. The other 26 objects
have both double-peaked H$\alpha$ and H$\beta$ emission lines {
(because we have a  limit on the redshift of dbp emitters selected
from SDSS less than 0.33).} This allows us a more convenient way to
measure and estimate the FeII emission lines as discussed below. The
objects are listed in Table 1.

\section{Results}
\subsection{Measurement of the line parameters of emission lines}

   Due to the presence of double-peaked broad H$\alpha$, in order to
get the best fitted results for narrow emission lines, there are
three to four broad gaussian functions for the broad components of
H$\alpha$, one single narrow gaussian function for each narrow
emission line in the wavelength range between 6250$\AA$ and
6850$\AA$ and a power law for the continuum. Thus, there are three
to four broad gaussian functions for broad H$\alpha$ and seven
narrow gaussian functions for [NII]$\lambda6548,6583\AA$,
[SII]$\lambda6719,6732\AA$, [OI]$\lambda6300,6363\AA$ and narrow
component of H$\alpha$. Here, we limit the center wavelength of each
doublet to the same redshift, and limit the velocity dispersion of
gaussian function of each doublet to the same value in velocity
space. Furthermore, we fix the flux ratio of [NII]$\lambda6548\AA$
to [NII]$\lambda6583\AA$ to 0.33, { limit the line width of narrow
emission line ($\sigma_{line}$) less than $400{\rm km\cdot s^{-1}}$
and limit the line width of broad emission lines ($\sigma{line}$) larger
than $500{\rm km\cdot s^{-1}}$}.
The last best fitted results are shown in Fig 1.

   According to the best fitted results for narrow emission lines, we can get
the broad components of H$\alpha$ in the wavelength range from
6250$\AA$ and 6850$\AA$. This provides a better way to measure the
FeII emission lines at optical bands and a better way to estimate
the model parameters of the accretion disk model.  Because of strong
FeII emission lines, it is much more difficult to obtain the
complete line profile of H$\beta$. However, we can estimate broad
double-peaked H$\beta$ by the line profile of H$\alpha$ after the
subtraction of narrow emission lines. Before proceeding further, we
should first subtract the continuum near H$\beta$. We subtract the
continuum according to the best fitted results for the points near
4100$\AA$ and near 5800$\AA$. The continuum near H$\beta$ is also
shown in Fig 1.

   Once we obtain the spectra in the wavelength range between 4100$\AA$ and
5800$\AA$ after the subtraction of the continuum, we can fit the spectra as
we have done in paper I, under the assumption that FeII emission lines come
from the same regions as broad Balmer emission lines:
\begin{equation}
\begin{split}
f_\lambda = &k_{H\beta}\times H\alpha + k_{H\gamma}\times H\alpha + k_{FeII}\sum FeII + \\
&k_{FeIII}\sum FeIII+ k_{MgI}\sum MgI + H\beta_{narrow} + \\
&H\gamma_{narrow} + [OIII]
\end{split}
\end{equation}
where  $k_{line}$ is the flux ratio of the emission line to broad
H$\alpha$. The narrow component of H$\beta$, H$\gamma$ and [OIII]
doublet can be fitted by narrow gaussian functions. If necessary,
there is a broad component with the same line profile as that of
broad H$\alpha$ for HeII$\lambda4686\AA$, and two extended gaussian
functions for the extend wings of the [OIII] doublet. The best
fitted results are shown in Fig 2. The parameters for the flux ratio
of different emission lines are listed in Table 2.

   { From the fitted results shown in Fig 2, we notice that there are
some objects for which the emission lines near H$\beta$, especially
near 4600$\AA$, can not be better fitted. The main reasons are
perhaps the following: First, we do not consider the contributions
of narrow emission lines of H$\gamma$+[OIII] and CVI emission lines,
since they are weak emission lines. The second reason  the accuracy
of the continuum subtraction. We fit the continuum under H$\beta$ by
one power law function. However, From the study of composite spectra
of quasars (Francis et al. 1991; Zheng et al. 1997; Berk et al.
2001), especially the study of composite spectra of AGN selected
from SDSS, we can see that the continuum should be best fitted by
two power laws with a break of $\sim5000\AA$. Thus, applying only
one power-law to fit the continuum under H$\beta$ should lead to
some uncertainty. }

   After the subtraction of narrow emission lines near H$\alpha$, we can
obtain the parameters of the elliptical accretion disk model
(Eracleous et al. 1995) { by the Levenberg-Marquardt technique
and adjusting the 8 free parameters (MPFIT package in IDL)}. There
are eight free parameters in the elliptical accretion disk model,
the inner radius $r_{in}$, the outer radius $r_{out}$, the
inclination angle of the accretion disk $i$, the local broadening
velocity dispersion $\sigma$, the slope of the line emissivity $q$,
the eccentricity $e$, the orientation angle of the accretion disk
$\phi_0$ and an amplitude factor $k$. The best fitted results are
shown in Fig 3. The parameters of the accretion disk model for each
dbp emitter are listed in Table 2.

   We select elliptical accretion disk model rather than circular disk
plus spiral arm model, due to the following considerations. For
elliptical disk model, there are mainly  four parameters to
determine: The observed line profile, $r_{in}$, $r_{out}$, $i$ and
$e$, however in the circular disk plus spiral arm model, there are
another more four parameters that dominate the observed line
profile. To some extend, we think the uniqueness of model parameters
for elliptical accretion disk model is better than circular disk
plus spiral arm model. { Also, we do not select a simple circular
accretion disk model, because the model can not be applied for dbp
emitters which have brighter red than blue peak.}

  Last, we measure the line parameters of broad H$\alpha$, such as the
line width of H$\alpha$, $\sigma_{H\alpha_B}$. The line parameters of
the broad component of H$\alpha$ can be measured by:
\begin{subequations}
\begin{align}
&\lambda_0 = \frac{\int\lambda f_{\lambda}d\lambda}{\int f_\lambda d\lambda} \\
&\sigma_g^2 = \frac{\int\lambda^2f_{\lambda}d\lambda}{\int f_\lambda d\lambda} - \lambda_0^2 = \frac{\int(\lambda-\lambda_0)^2f_{\lambda}d\lambda}{\int f_\lambda d\lambda}
\end{align}
\end{subequations}
We find that the mean value of $f_{\lambda}$ after the subtraction
of the continuum is zero for the nearby points on the continuum,
however, the mean value $(\lambda-\lambda_0)^2f_{\lambda}$ for those
points further out is not zero, and becomes larger for the points
selected from more extend distance from the center of the emission
line, unless we use a  smoother background, after the subtraction of
the continuum, for which the value of each point is equal to zero.
Thus, we should select cautiously the wavelength to measure the
second moment according to the equation above. The wavelength range
is selected as follows: after the subtraction of the continuum and
the bad pixels, we select the last point for  which the flux is
equal to the minimum value (sometimes, the minimum value of the flux
is not equal to zero) of the points on the blue side of the emission
line and select the first point for which the flux is equal to the
minimum value of the points on the red side of the emission line.
Due to the composite nature of the line profile of double-peaked
emission, we think it is more accurate to measure the second moment
of the broad H$\alpha$ using the equation above, than to measure the
value of FWHM of H$\alpha$ according to the standard definition of
FWHM, because the flux density of the broad component is the largest
one in the composite emission line after the subtraction of the
narrow emission lines. The value of $\sigma_{H\alpha_B}$ is listed
in Table 1.

\subsection{Properties of FeII emission lines of dbp emitters}

   First, we check the correlation between line width of broad Balmer
emission lines and strength of FeII emission lines
EW(FeII)/EW(H$\beta$). We show the correlation in Fig 4.
{ The mean value of the second moment of broad H$\alpha$ is about
$3002\pm139{\rm km\cdot s^{-1}}$}. As pointed out above, we use the second
moment of broad H$\alpha$ rather than FWHM(H$\alpha$).
{ The Spearman Rank correlation coefficient is -0.39 with
$P_{null}\sim5\%$.} Here, the data include the object
SDSS J2125-0813, for which the second moment of broad H$\alpha$ is
estimated from the broad component of H$\beta$ reproduced by the
accretion disk model. The results indicate that the anti-correlation
between EW(FeII)/EW(H$\beta$) and line width of broad lines is also
valid for dbp emitters.

   The correlation between FeII properties and the line width of
Balmer broad lines reflects, to some extent, effects of the
line-of-sight. In the case of the correlation between the FeII
properties and the width of the narrow lines, which are considered
as tracers of stellar velocity dispersion, the effect of the
bulge-BH mass on the properties of FeII emission may be more neatly
deconvolved. A detailed description about this is shown in the next
subsection. Fig 5 shows the correlation between
EW(FeII)/EW(H$\beta$) and line width of narrow lines. Here, we
select the line width of the [OIII] emission line. As described
above, there two gaussian components are needed for [OIII] emission
line: one normal (core) component and an extended one. We use the
core components as described in Greene \& Ho (2005a). The Spearman
Rank of the correlation coefficient is -0.52 with
$P_{null}\sim0.6\%$.

   Furthermore, we check the radio properties of dbp emitters with
apparent FeII emission lines.
We find that there are 8 dbp emitters in our sample targeted and 13
dbp emitters covered by FIRST/NVSS. Thus, we can
get the radio flux density or the upper limit value of radio flux density
at 20cm. According to the definition of
radio loudness, $R_i$, by Ivezi$\rm{\acute{c}}$ et al. (2002):
\begin{equation}
R_{i}=log(F_{20cm}/F_{i})=0.4\times(m_i-t)
\end{equation}
where $m_i$ is one of the SDSS magnitudes and t is 20cm AB radio
magnitude defined as $t = -2.5\cdot\log(F_{20cm}/3631\rm{Jy})$,
we can calculate radio loudness for each dbp emitter in our sample.
There are only one dbp emitter with $R_r>1$, SDSS J075407.95+431610.5,
which is one radio loud quasar. The other 20 dbp emitters are radio quiet
quasars. Thus dbp emitters with stronger FeII emission lines are apt to
be found in radio quiet quasars, which in agreement with that of E1 analysis
of single peaked BRL AGN (e. g. Sulentic et al. 2003).

 It is also convenient to check the
correlation between the continuum luminosity and the luminosity of
H$\alpha$ which is confirmed by Greene \& Ho (2005b) using a sample
of AGN selected from SDSS, because in our code, the flux of
H$\alpha$ is one of fundamental parameters to determine the flux of
other emission lines, such as the flux of double-peaked H$\beta$,
$k_{H\beta}\times f_{H\alpha}$, the flux of FeII emission lines,
$K_{FeII}\times f_{H\alpha}$.
One of the main considerations is that the correlation reflects
some physical parameters of BLRs of AGN, such as the covering factor
of BLRs, the proportional contribution of the ionization energy to
the emission of broad emission lines etc..
The correlation is shown in Fig 7. The
Spearman Rank correlation coefficient is $\sim1$ with
$P_{null}\sim0$. The correlation for dbp emitters
obeys the same relation as that for normal AGN,
$L_{H\alpha}\sim L_{5100\AA}^{1.157}$.
{ Here, the luminosity of H$\alpha$ also includes
the contributions of [NII] doublets, besides the contributions
of broad and narrow components of H$\alpha$, because sometimes
it is difficult to separate the narrow component of H$\alpha$
from the components of [NII] for some dbp emitters.
Because of the smaller contributions of [NII] emission
lines, it is not affect the result between $L_{H\alpha}$ and
$L_{5100\AA}$.}

   In the case of dbp we can get, from the best fitted results of the
accretion disk model, the inclination angle of the accretion disk.
The plot of the strength of FeII \emph{vs.} the inclination angle of
the accretion disk is shown in Fig 8.
{ The Spearman Rank correlation
coefficient is 0.43 with $P_{null}\sim6\%$. We
selected only 20 objects with accurate inclination angles $i$,
$i>1.5\times i_{error}$.} The inclination angle derived from the
accretion disk model has a somewhat large error, specially when the
angle is near 90 degrees.
{ We cannot confirm a trend for all objects,
because, if we only consider the objects with inclination angle less
than 60 degrees, the coefficient is 0.16 with $P_{null}\sim63\%$.}
On the other hand, if the trend between the strength of FeII
emission lines and the inclination angle shown in Fig. 7 was true,
it would indicate that the objects with broader Balmer lines have
stronger FeII emission lines, which is in contradiction with the
much stronger correlation between strength of FeII emission lines
and the line width of broad H$\alpha$. And in contradiction with all
previous results found from E1 parameter space analysis, e.g.
Sulentic, Marziani \& Dultzin-Hacyan (2000). Thus, the trend most be
a fake one. This is due not only to the fact that values for the
inclination have large errors for large angles, but also due to the
fact that inclination alone cannot govern the FeII strength, it is
always convolved with BH mass and accretion rate, as discussed below
(see also Marziani et. al. 2003a)

\subsection{Influence of Black Hole mass and Accretion Rate}

  The most accurate way to estimate the central BH masses is by the
stellar velocity dispersion as in Gebhardt et al. (2000); Ferrarese
\& Merritt (2001); Tremaine et al. (2002).
For quasars, it is not possible to determine the BH masses by
stellar velocity dispersion. Another convenient way
is under the assumption of Virialization (Greene \& Ho 2005b; Ovcharov et al.
2005; Wu et al. 2004; McLure \& Jarvis 2004; 2002; Marziani et
al. 2006; Onken et al. 2004; Peterson et al. 2004).
However, there are various caveats for the
measurement of the full width at half maximum (FHHM) of broad
low-ionization  emission lines, such H$\beta$, especially for
dbp emitters. A detailed discussion
of the line components that do reverberate can be found  in Sulentic
et al. (2006). Actually, these authors prefer to use the FeII lines
(assuming they are virialized as we prove here to determine BH
masses). Unfortunately, for double-peaked broad emission lines, it is
not reasonable to fit the line by one gaussian function or to
measure the value of FWHM for double-peaked emission lines.

    However, there is a strong correlation
between stellar velocity dispersion and line width of low-ionization
narrow emission lines, which has been studied by many authors for different
kinds of AGN. Nelson \& Whittle (1996) found that there is a
correlation between line width of [OIII]$\lambda5007\AA$ and stellar
velocity dispersion. Wang \& Lu (2001) also emphasized the relation
for a sample of NLS1s. Recently, Greene \& Ho (2005a) confirmed that
the line width of low-ionization narrow emission lines can trace the
stellar velocity dispersion using a sample of low luminous AGN
selected from SDSS (York et al. 2000; Strauss et al. 2002;
Abazajian et al. 2004). Zhou et al. (2006) re-confirmed the
stronger relation for a larger sample of NLSy1 selected from SDSS by
line width of [NII] emission line. As a test, Boroson (2003) have
shown that the coincidence between the BH masses estimated from the
relation $M_{BH} - \sigma_{[OIII]}^{4.02}$ and the BH masses
estimated under the assumption of virialization for a sample of AGN
selected from SDSS.

  As mentioned above, in order to determine the BH masses of the
dbp emitters, we use the line width of [OIII] emission line as the
estimator, according to the following equation (Gebhardt et al. 2000;
Ferrarese \& Merritt 2001; Tremaine et al. 2002):
\begin{equation}
M_{BH} = 10^{8.13\pm0.06}(\frac{\sigma_{[OIII]}}{200{\rm km\cdot s^{-1}}})^{4.02\pm0.32} {\rm M_{\odot}}
\end{equation}
where $\sigma_{[OIII]}$ is the line width of normal component of
[OIII]$\lambda5007\AA$.
The correlation between BH masses and EW(FeII)/EW(H$\beta$) has been
pointed out above as the correlation between
$\sigma_{[OIII]\lambda5007\AA}$ and EW(FeII)/EW(H$\beta$) (Fig. 5)
The anti-correlation of
EW(FeII)/EW(H$\beta$) and line width of [OIII]$\lambda5007\AA$
indicates that the dbp emitters with lower BH masses have stronger
FeII emissions. This result is in agreement with that of E1 analysis
of single peaked BRL AGN (e. g. Sulentic, Marziani \& Dultzin-Hacyan
2000).

  Here we compare with the correlation between the dimensionless
accretion rate and EW(FeII)/EW(H$\beta$) shown in Fig 9. The
dimensionless accretion rate $\dot{m}$ is calculated as
$\dot{m}=\frac{9\times L_{5100\AA}}{M_{BH}\times1.38\times10^{38}}$
(Wandel et al. 1999; Kaspi et al. 2000). The Spearman Rank
correlation coefficient is 0.23 with $P_{null}\sim24\%$.
Such a weak correlation indicates that dimensionless accretion rate
has weaker effects on FeII emission of dbp emitters than BH mass.

   The mean BH masses for the 27 dbp emitters is
$10^{8.05\pm0.11}{\rm M_{\odot}}$, the mean dimensionless accretion rate
is $10^{-1.04\pm0.11}$. According to the dimensionless accretion
rate, the accretion mode in the central region is standard accretion
rather than ADAF mode, because the upper limit accretion rate for
ADAF mode is about $\dot{m}\sim0.28\times\alpha^2$ where $\alpha$ is
viscous coefficient $\alpha\sim0.1-0.3$ (Mahadevan \& Quataert,
1997; Lasota et al., 1996; Narayan et al., 1995). If we select
$\alpha\sim0.1$ (Yi, 1996; Choi et al, 2001), all the dbp emitters
in our sample have larger accretion rate than the critical value 0.0028.
Furthermore,
the 27 dbp emitters have normal quasar spectra, which have normal
big blue bumps. Thus, even if ADAF accretion mode dominates the
accretion flow in the inner region of the accretion disk, especially
for the two objects with lower accretion rate, the standard
accretion mode should mainly dominate the bulk of accretion.

\section{Discussions and Conclusions}

  We have accurately reproduced the line profiles in the wavelength range from
4100$\AA$ to 5600$\AA$ by the line profile of double-peaked broad
H$\alpha$. The best fitted results indicate that the broad optical
FeII emission lines are also double peaked and originate from the
same place where the double-peaked broad Balmer emission lines. More
and more evidence confirms that the double-peaked broad
low-ionization  lines come from the accretion disk rather than an
outflow or other models. In our sample, the double-peaked broad
H$\alpha$ of all the 27 dbp emitters can be best fitted by the
elliptical accretion disk model (Eracleous et al. 1995). The best
fitted results for optical FeII emission lines indicate that FeII
emission lines also originate from the accretion disk which provides
the needed high density for FeII emission lines.

  The accretion rate and the normal quasar spectra indicate that ADAF
accretion mode in the dbp emitters with FeII emission lines is not
the main accretion mode. BH masses have more influence on the FeII
properties of dbp emitters than dimensionless accretion rate. The
reliable anti-correlation between EW(FeII)/EW(H$\beta$) and
$\sigma_{H\alpha}$ indicate that for dbp emitters, the emission
region of stronger FeII lines are far from the center black hole. We
have shown that even for dbp emitters, the influence of orientation
on FeII intensity cannot be deconvolved from the other determining
physical parameters of these AGN: mainly BH mass and, to a lesser
extent, accretion rate.

  We also get the same strong correlation between
$L_{5100\AA}$ and $L_{H\alpha}$ (includes the broad and narrow
components near H$\alpha$) for dbp emitters in our ample as that
obeyed by normal AGN, which indicates that the connection between
the continuum and BLRs is the same for normal AGN and for dbp
emitters. We know the luminosity of broad emission lines, such as
broad H$\alpha$, can be estimated by $L_{H\alpha} =
h\nu_{H\alpha}n_en_H\alpha_{H\alpha}\epsilon V$, where,
$\alpha_{H\alpha}$ is the recombination coefficient for H$\alpha$,
$\epsilon V$ is the volume filled by gases in BLRs. The electron
density in the accretion disk is larger than that in the normal
BLRs, which indicates that the value of
$n_en_H\alpha_{H\alpha}\epsilon V$ is a constant value for AGN,
normal AGN and dbp emitters

{ Finally we want to stress that the correlations, or more
precisely perhaps, the trends pointed out above remain to be
confirmed, because of two main reasons: First, the comparative
numbers of well defined dbp emitters is small (which is intriguing
enough), and second, because one of our selection criteria (the
apparent strehgth of optical FeII) may bias the trends.}

\section*{Acknowledgements}
ZXG gratefully acknowledges the postdoctoral scholarships offered by
la Universidad Nacional Autonoma de Mexico (UNAM). D. D-H
acknowledges support from grant IN100703 from DGAPA, UNAM. This
research has made use of the NASA/IPAC Extragalactic Database (NED)
which is operated by the Jet Propulsion Laboratory, California
Institute of Technology, under contract with the National
Aeronautics and Space Administration. This paper has also made use
of the data from the SDSS projects. Funding for the creation and the
distribution of the SDSS Archive has been provided by the Alfred P.
Sloan Foundation, the Participating Institutions, the National
Aeronautics and Space Administration, the National Science
Foundation, the U.S. Department of Energy, the Japanese
Monbukagakusho, and the Max Planck Society. The SDSS is managed by
the Astrophysical Research Consortium (ARC) for the Participating
Institutions. The Participating Institutions are The University of
Chicago, Fermilab, the Institute for Advanced Study, the Japan
Participation Group, The Johns Hopkins University, Los Alamos
National Laboratory, the Max-Planck-Institute for Astronomy (MPIA),
the Max-Planck-Institute for Astrophysics (MPA), New Mexico State
University, Princeton University, the United States Naval
Observatory, and the University of Washington.

\newpage
\onecolumn
\begin{figure}
\includegraphics[height=5cm,width=84mm]{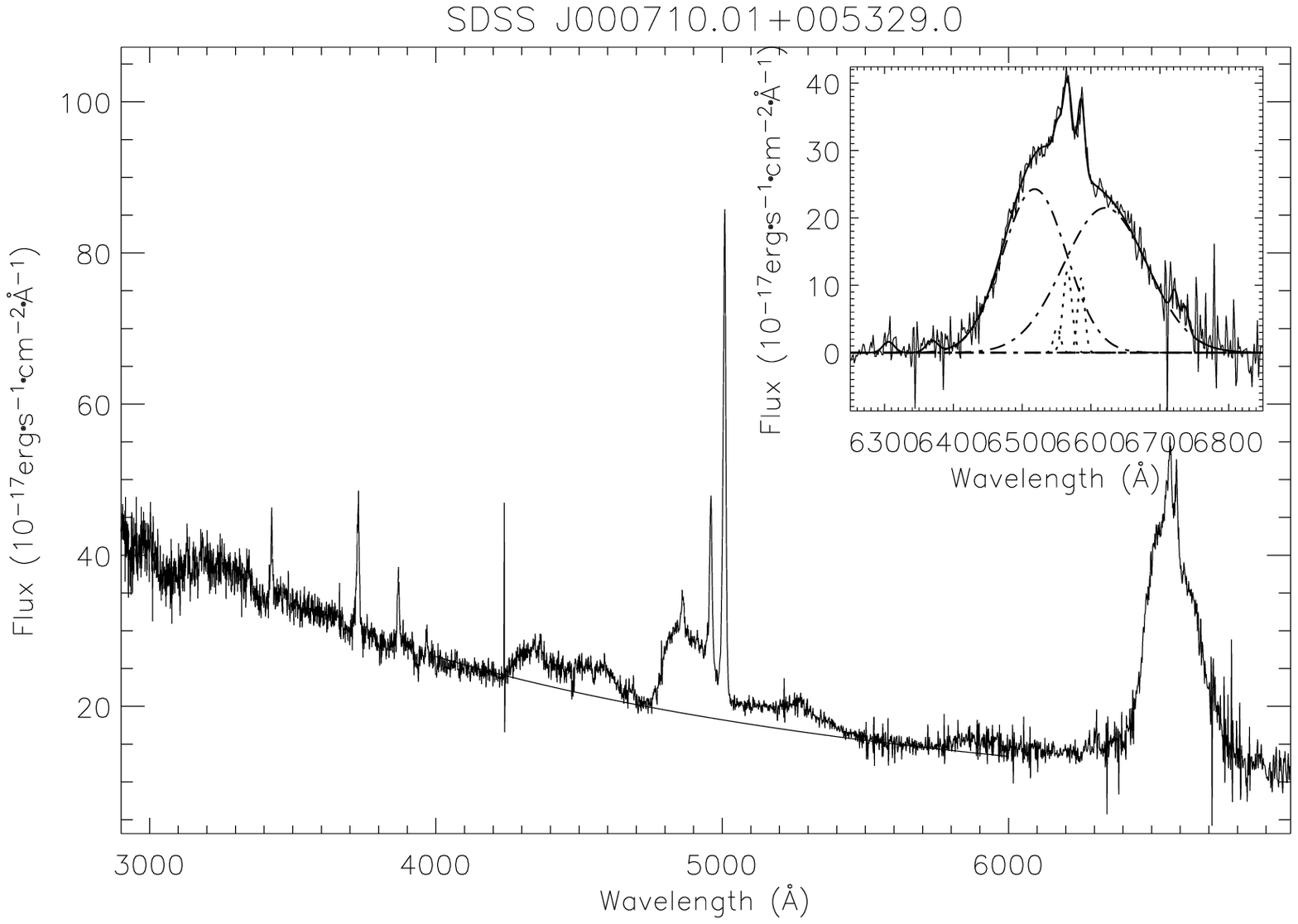}
\includegraphics[height=5cm,width=84mm]{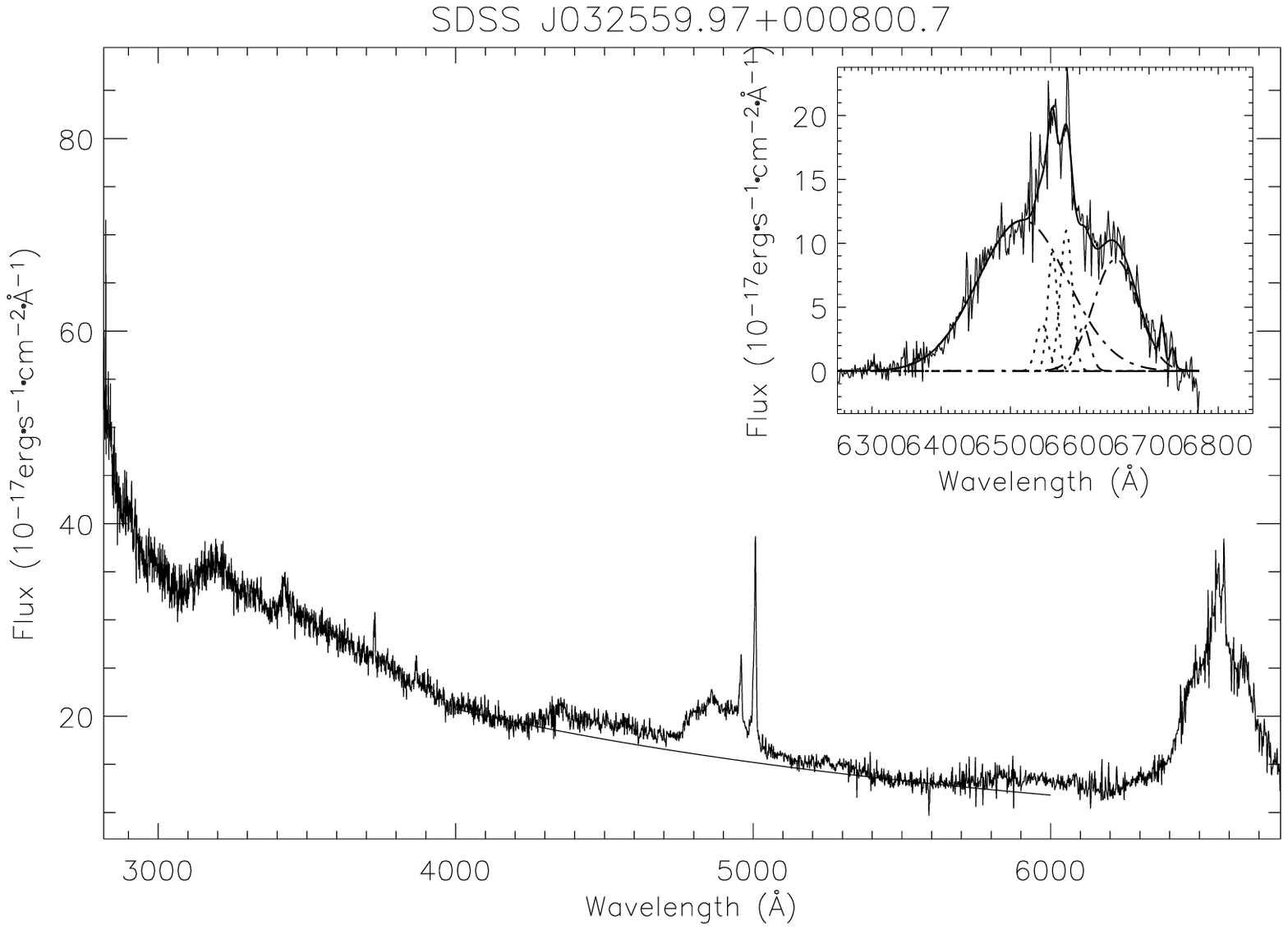}
\includegraphics[height=5cm,width=84mm]{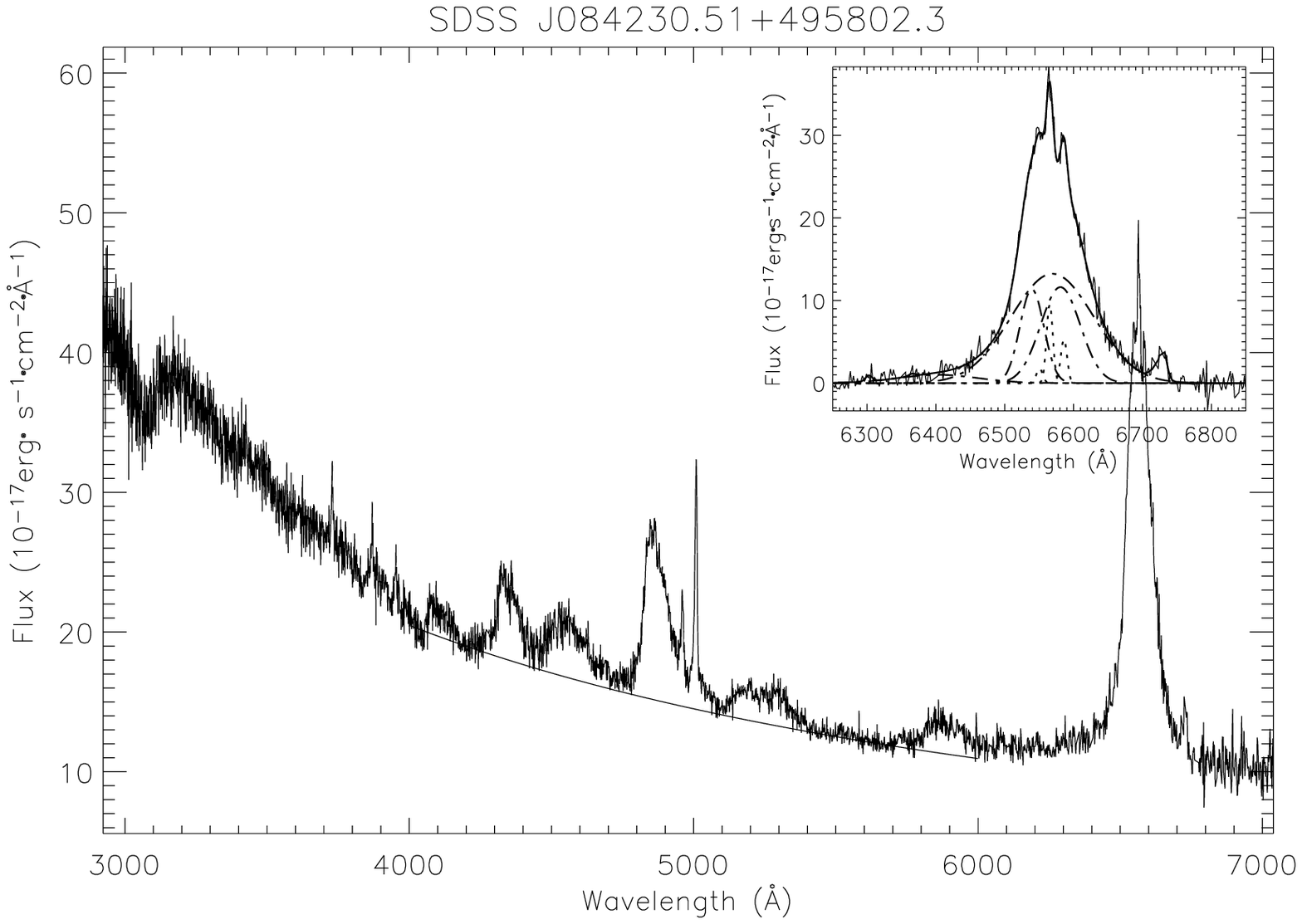}
\includegraphics[height=5cm,width=84mm]{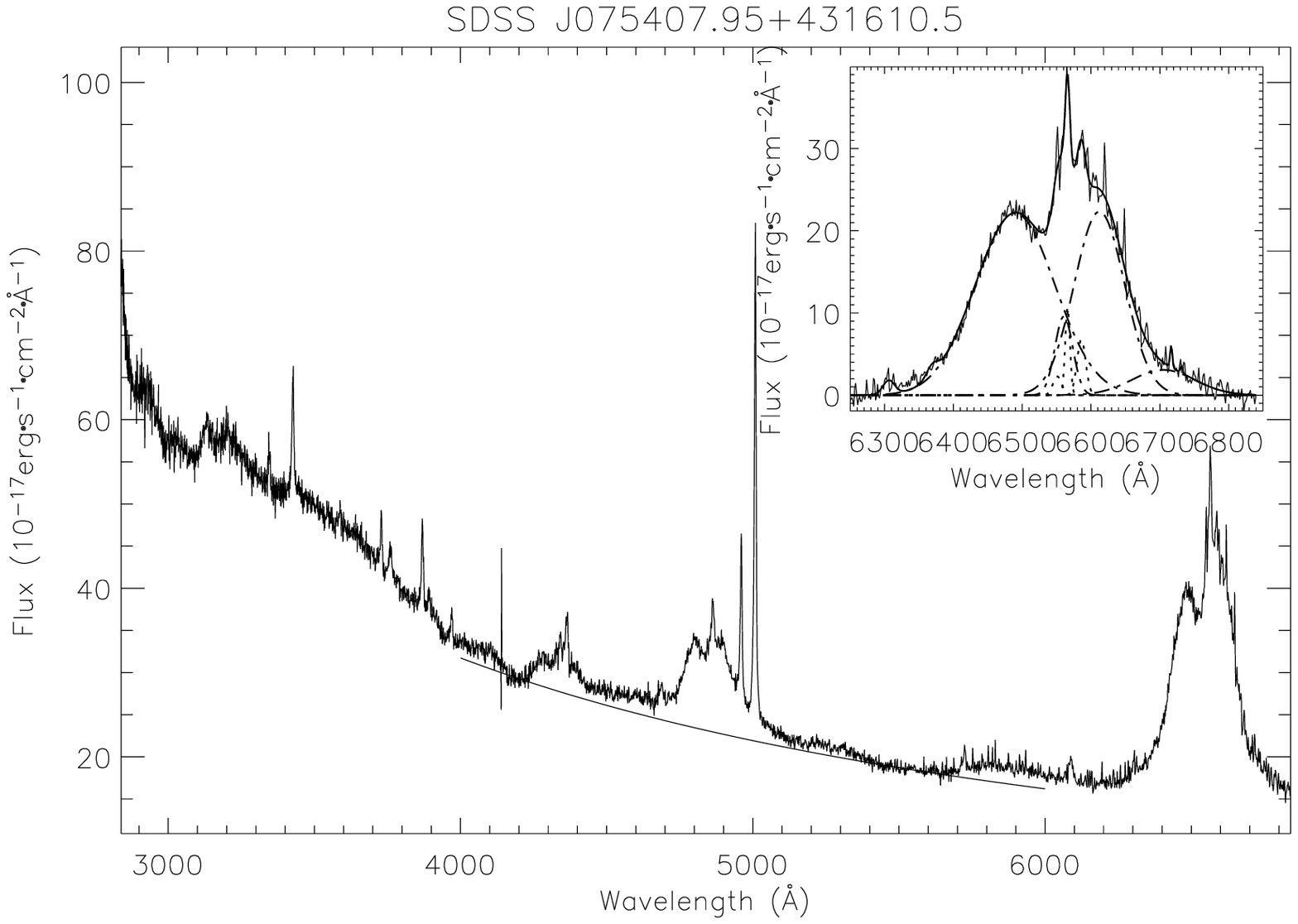}
\includegraphics[height=5cm,width=84mm]{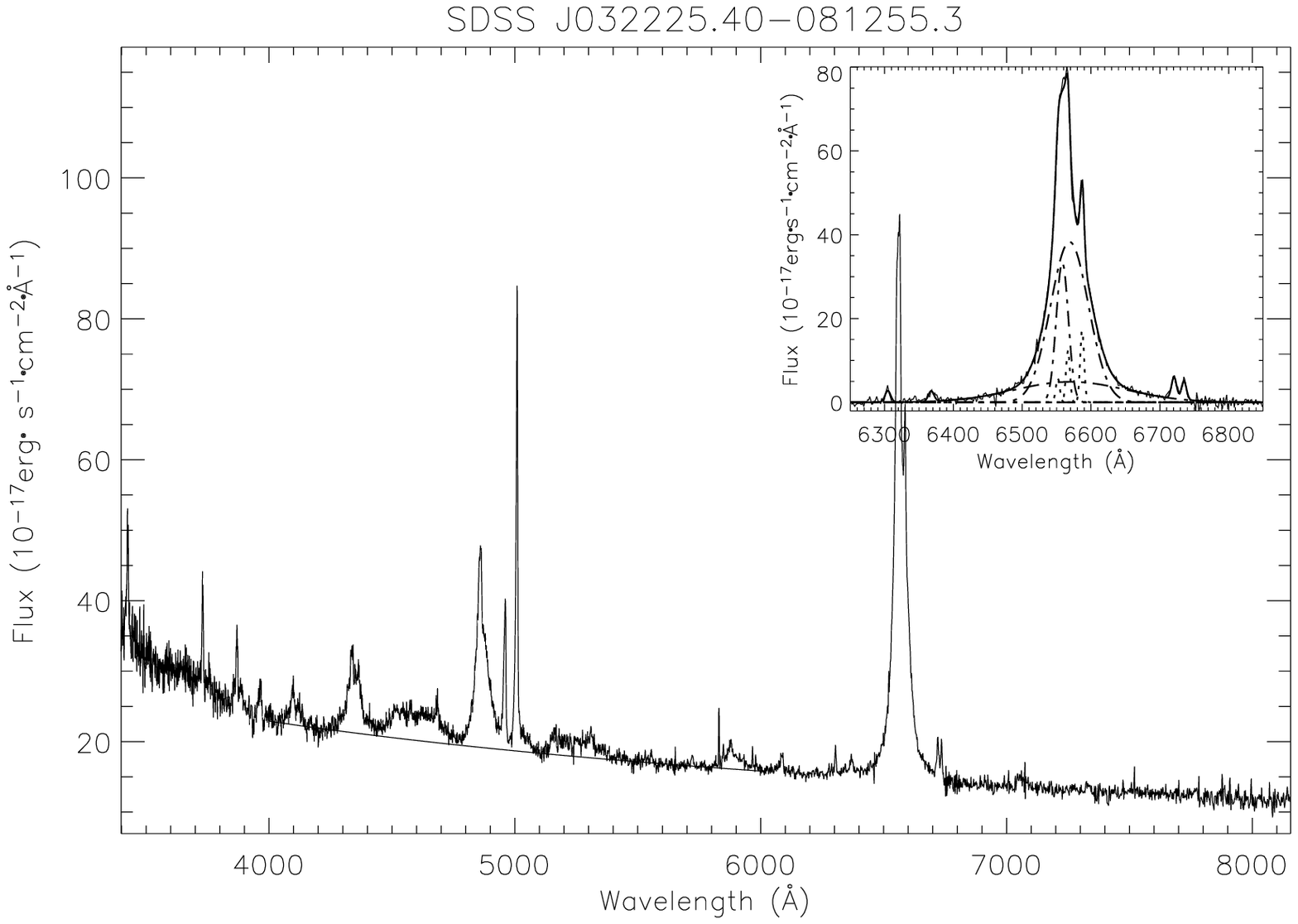}
\includegraphics[height=5cm,width=84mm]{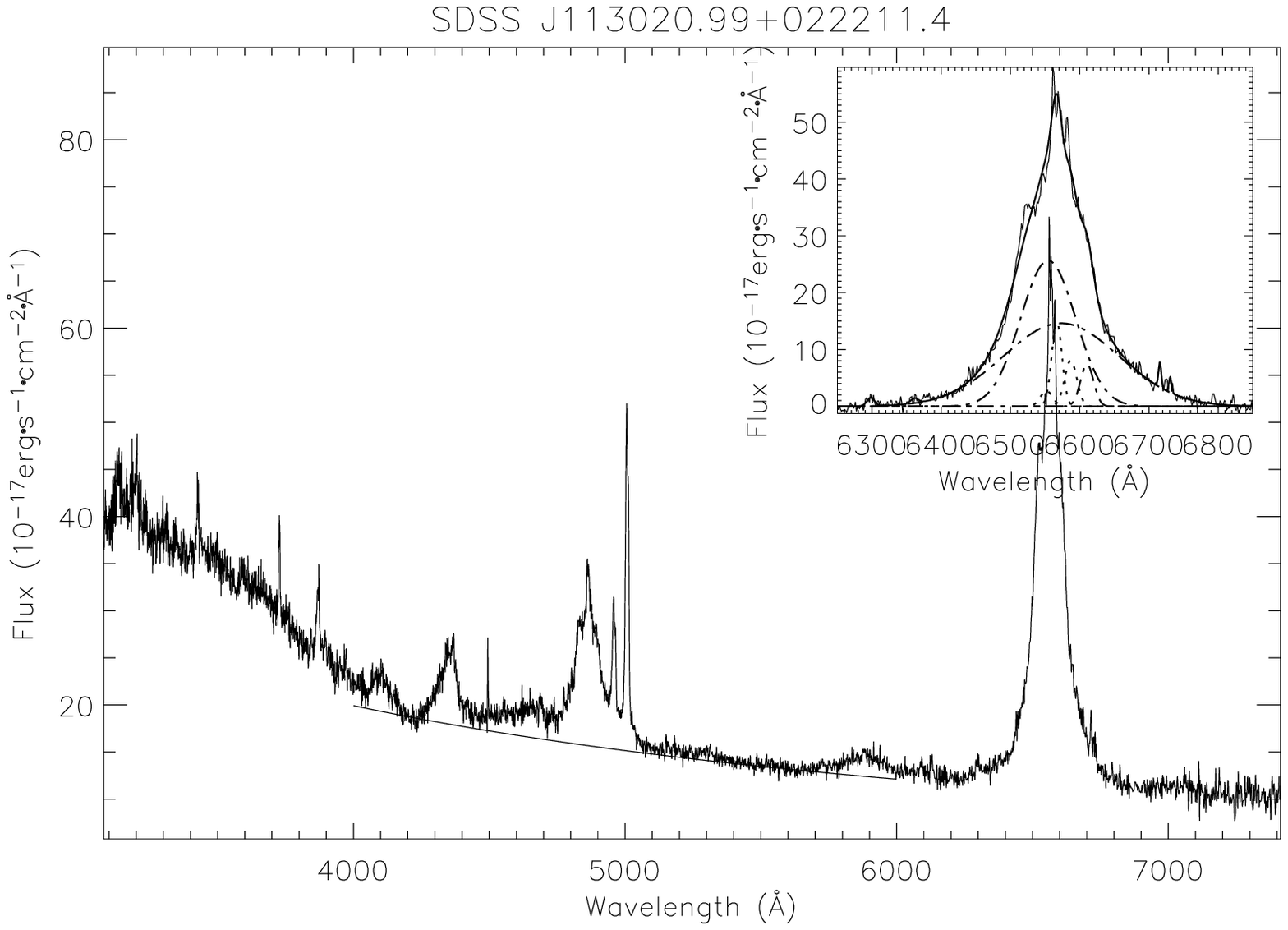}
\includegraphics[height=5cm,width=84mm]{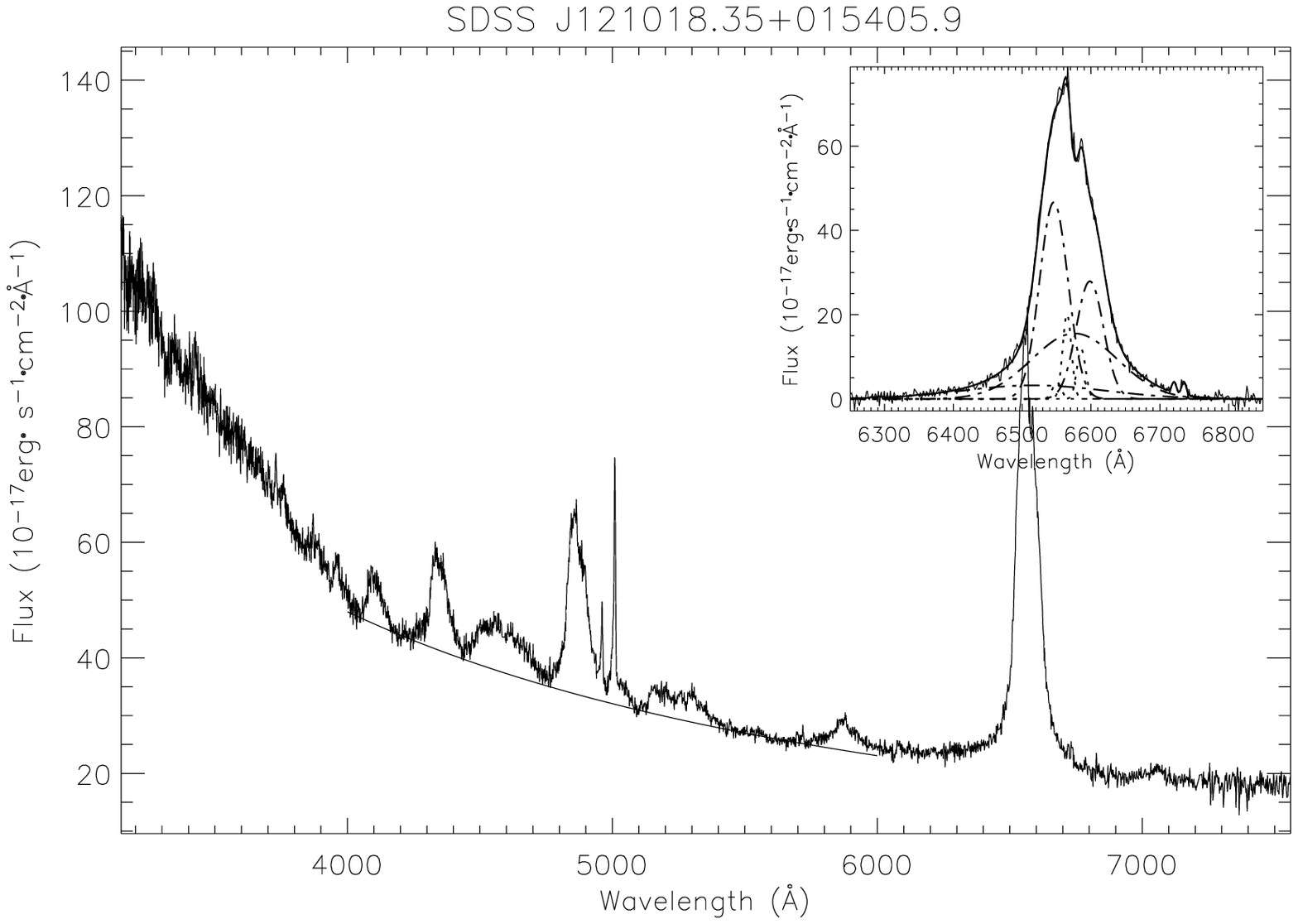}
\includegraphics[height=5cm,width=84mm]{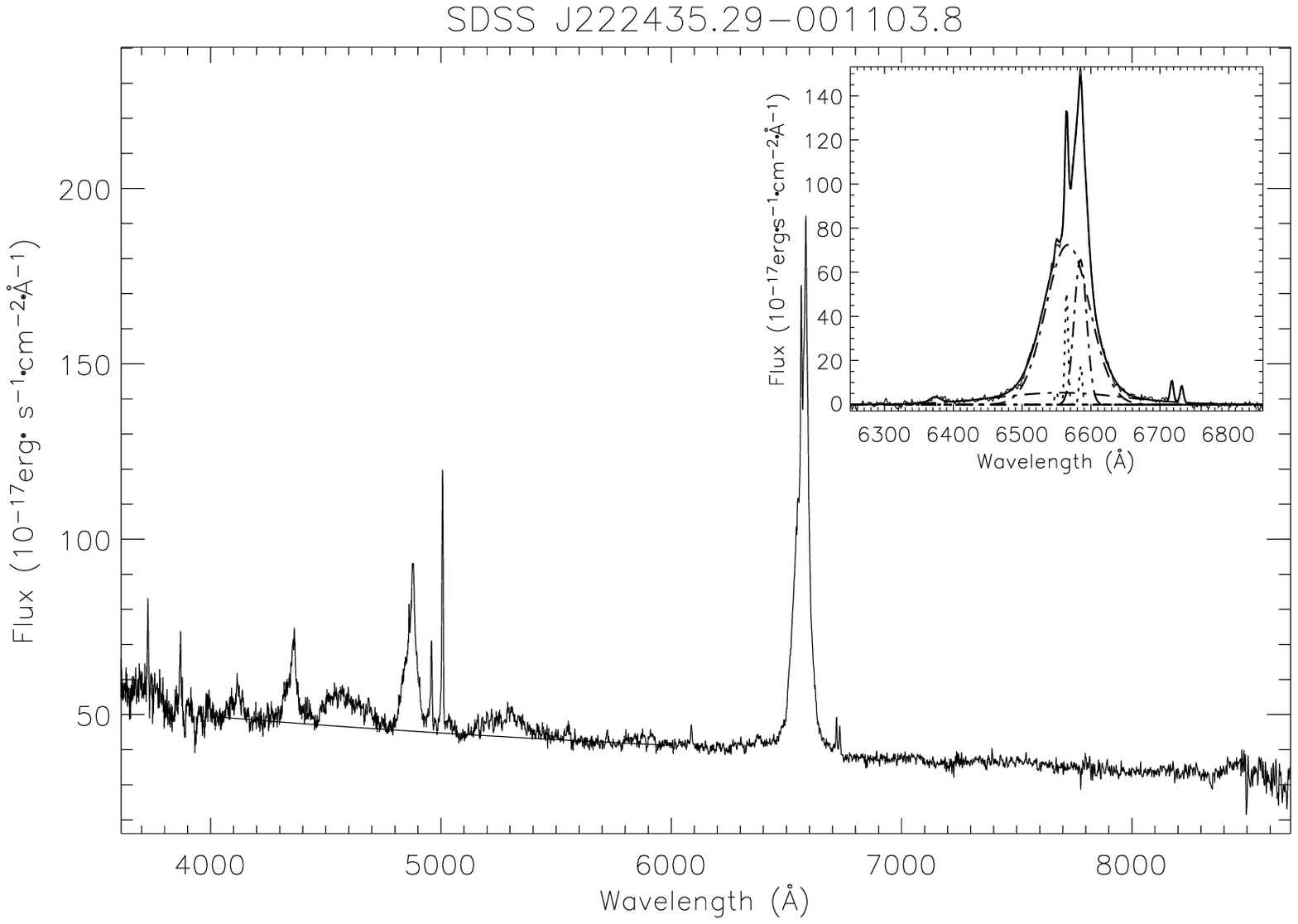}
\caption{The spectra of dbp emitters and the continuum under
H$\beta$ are shown in the figure. The best fitted results for narrow
emission lines near H$\alpha$ are shown in the upper right panels in
each plot. The thin solid line is the spectra after the subtraction
of the continuum. The thick solid line represents the best fitted
results, the dotted line presents the components for
[NII]$\lambda6548,6583\AA+$H$\alpha_{narrow}$. The dash-dotted line
represents the components for broad H$\alpha$. Here, only the first eight 
objects are shown in the figure.}
\end{figure}

\newpage
\onecolumn
\begin{figure}
\includegraphics[height=5cm,width=84mm]{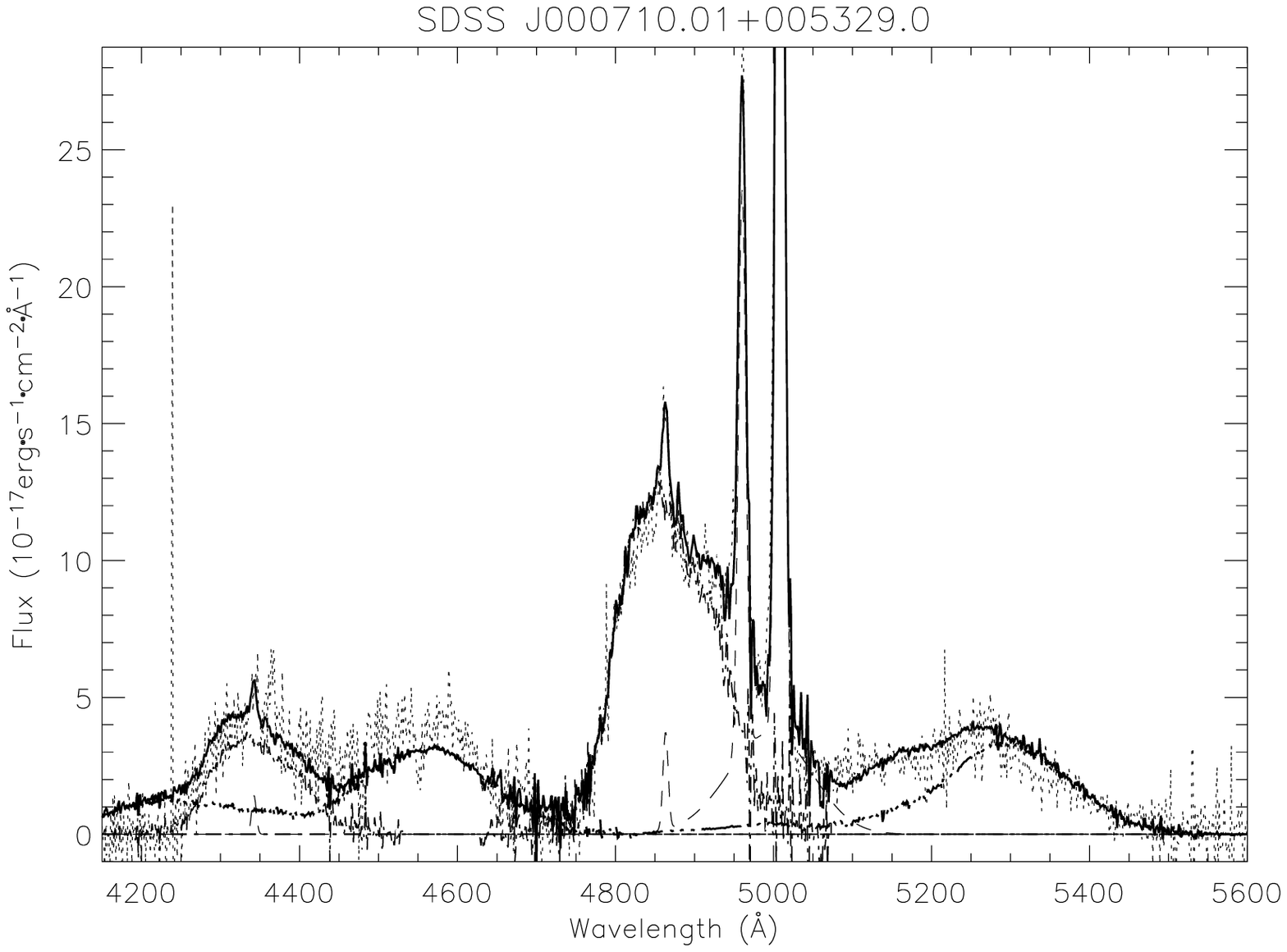}
\includegraphics[height=5cm,width=84mm]{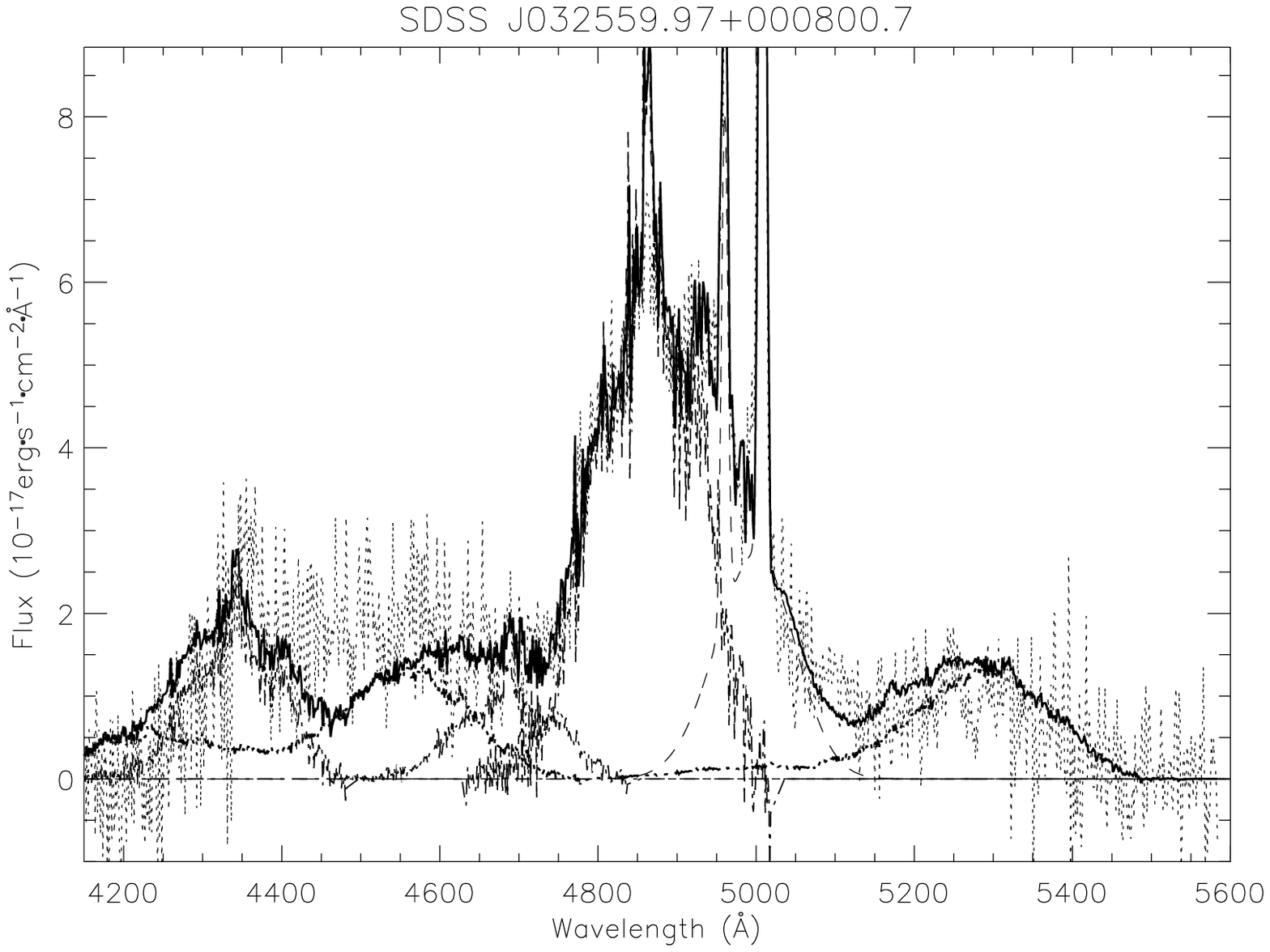}
\includegraphics[height=5cm,width=84mm]{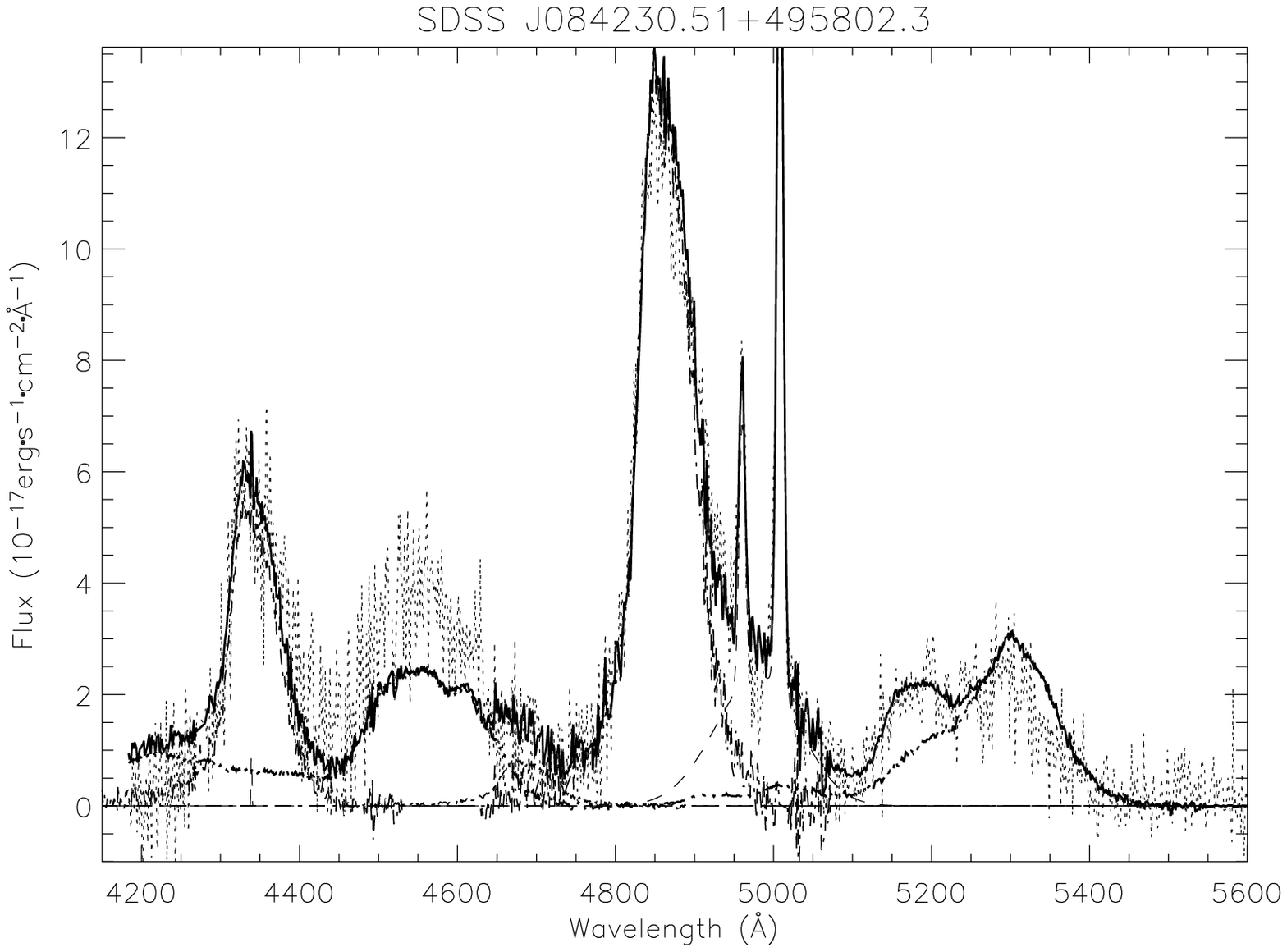}
\includegraphics[height=5cm,width=84mm]{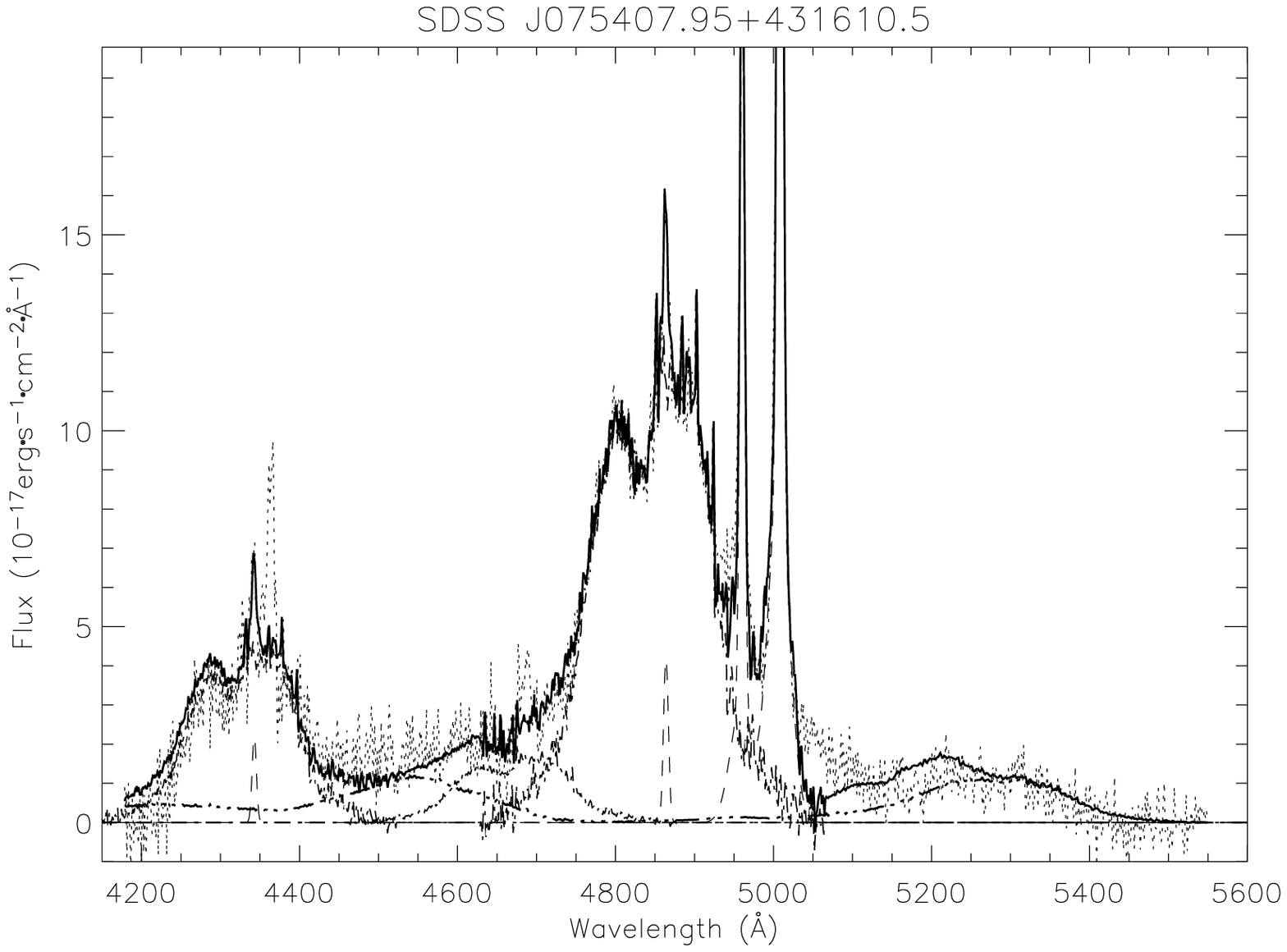}
\includegraphics[height=5cm,width=84mm]{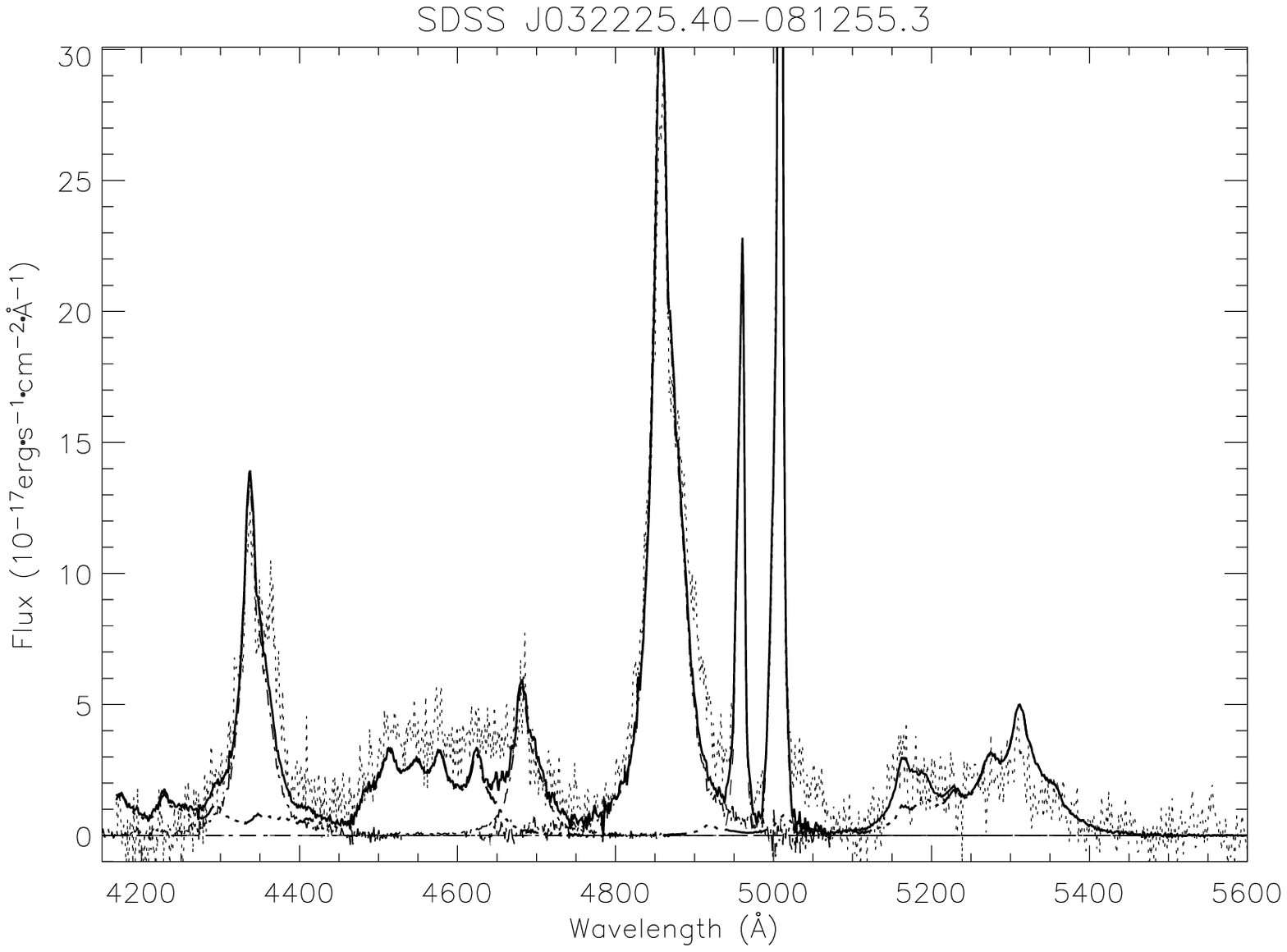}
\includegraphics[height=5cm,width=84mm]{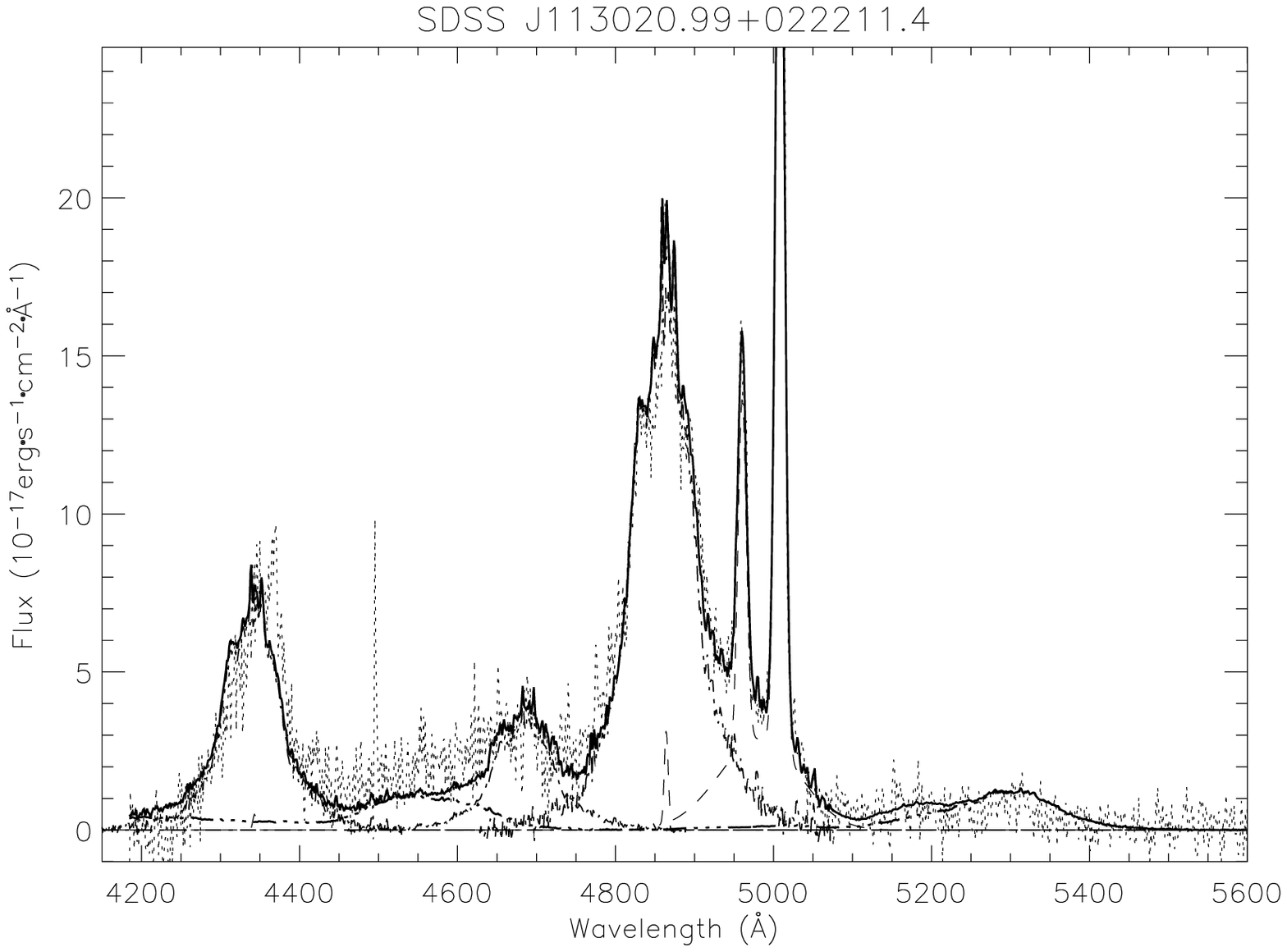}
\includegraphics[height=5cm,width=84mm]{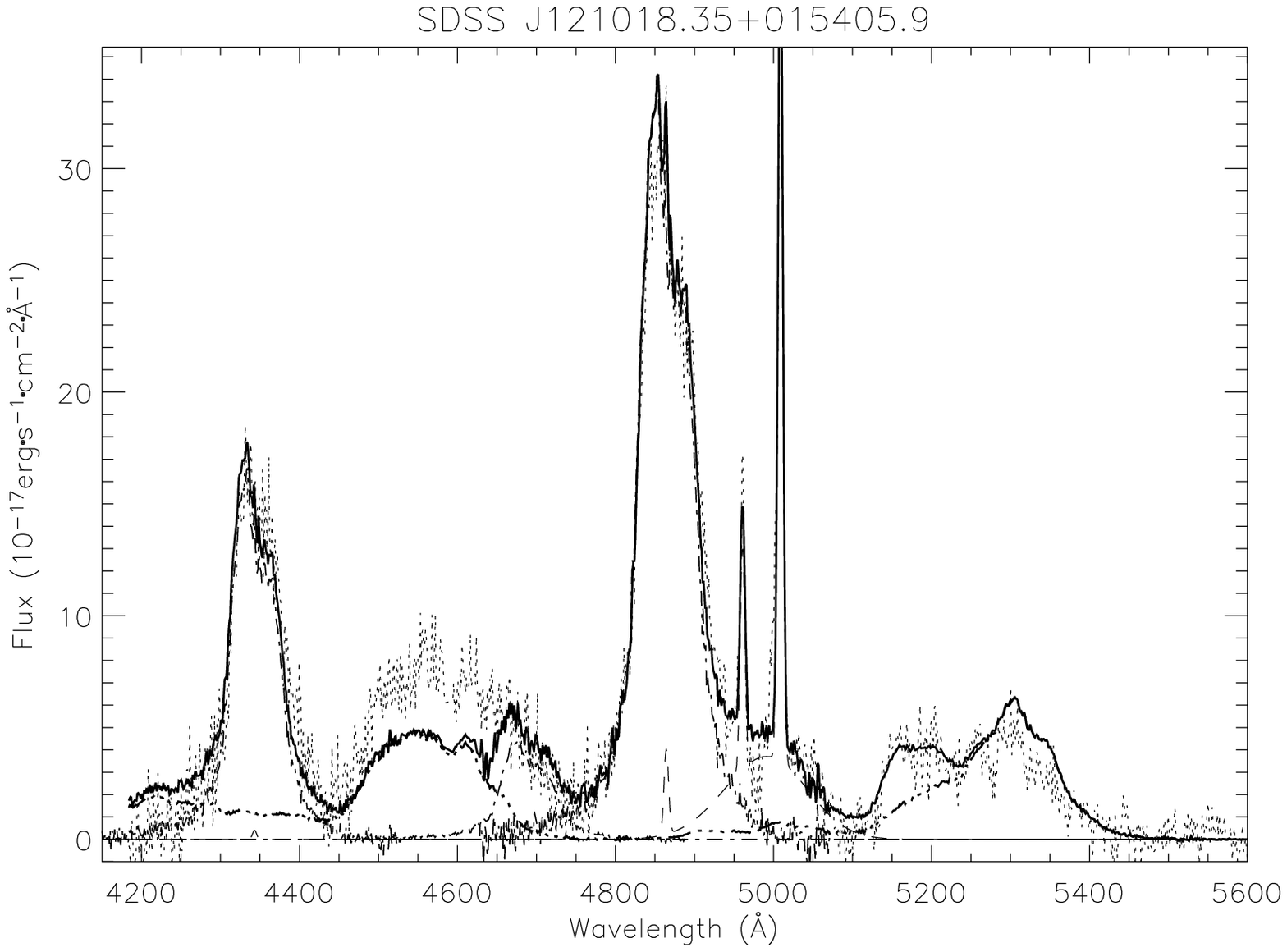}
\includegraphics[height=5cm,width=84mm]{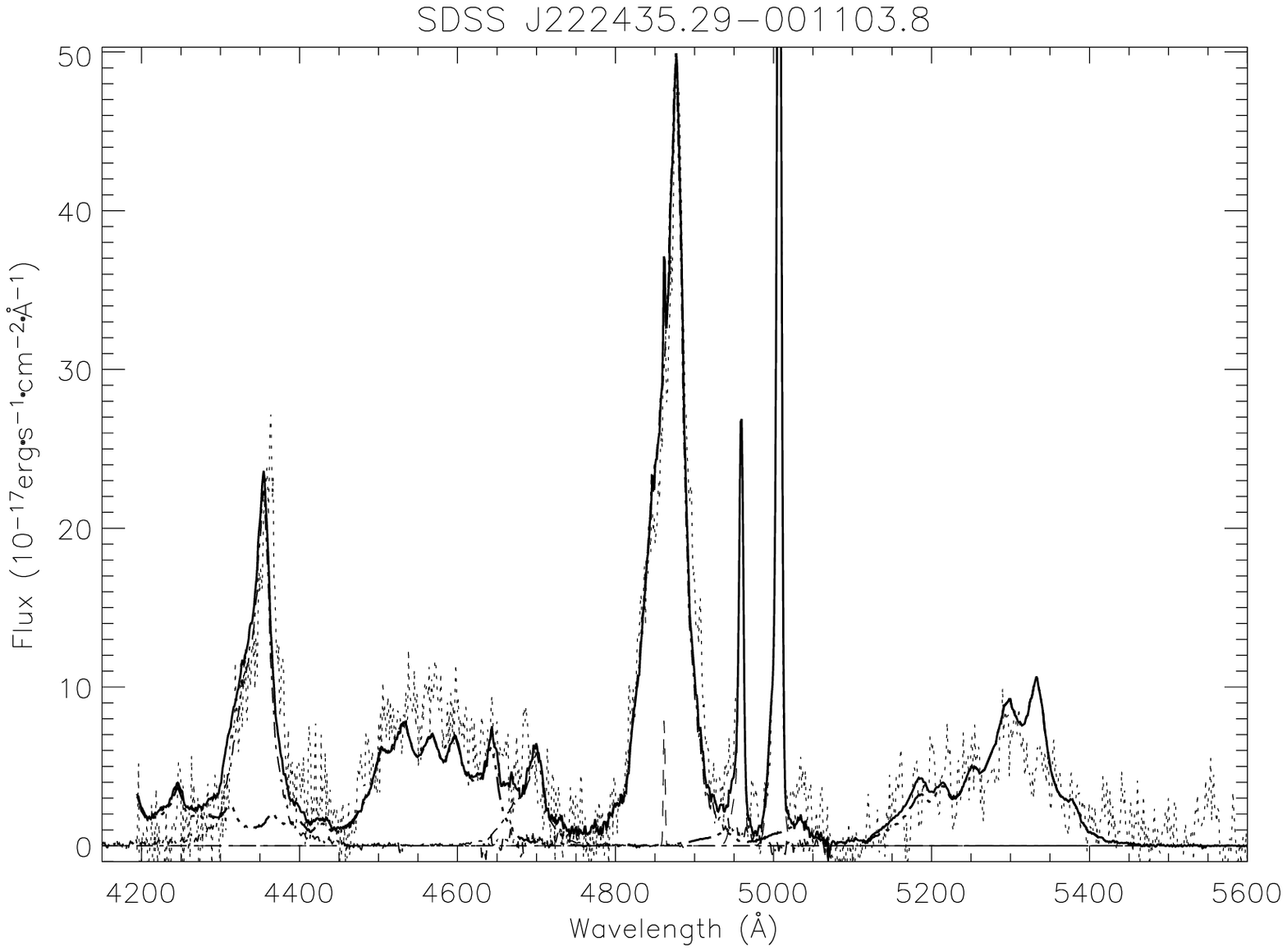}
\caption{The best fitted results for the spectra near H$\beta$ of dbp
emitters. The dotted line represents the line spectra after the subtraction
of the continuum. The thick solid line represents the best fitted results.
The dashed line represents the narrow components for [OIII] doublet and/or
H$\beta$ and/or H$\gamma$. The dot-dashed line represents the broad
components for H$\beta$, H$\gamma$ and/or HeII$\lambda4686\AA$. The double
dot-dashed line represents the components for FeII emission lines in the
wavelength range from 4200$\AA$ to 5600$\AA$.
Here, only the first eight
objects are shown in the figure.}
\end{figure}

\newpage
\onecolumn
\begin{figure}
\includegraphics[height=5cm,width=84mm]{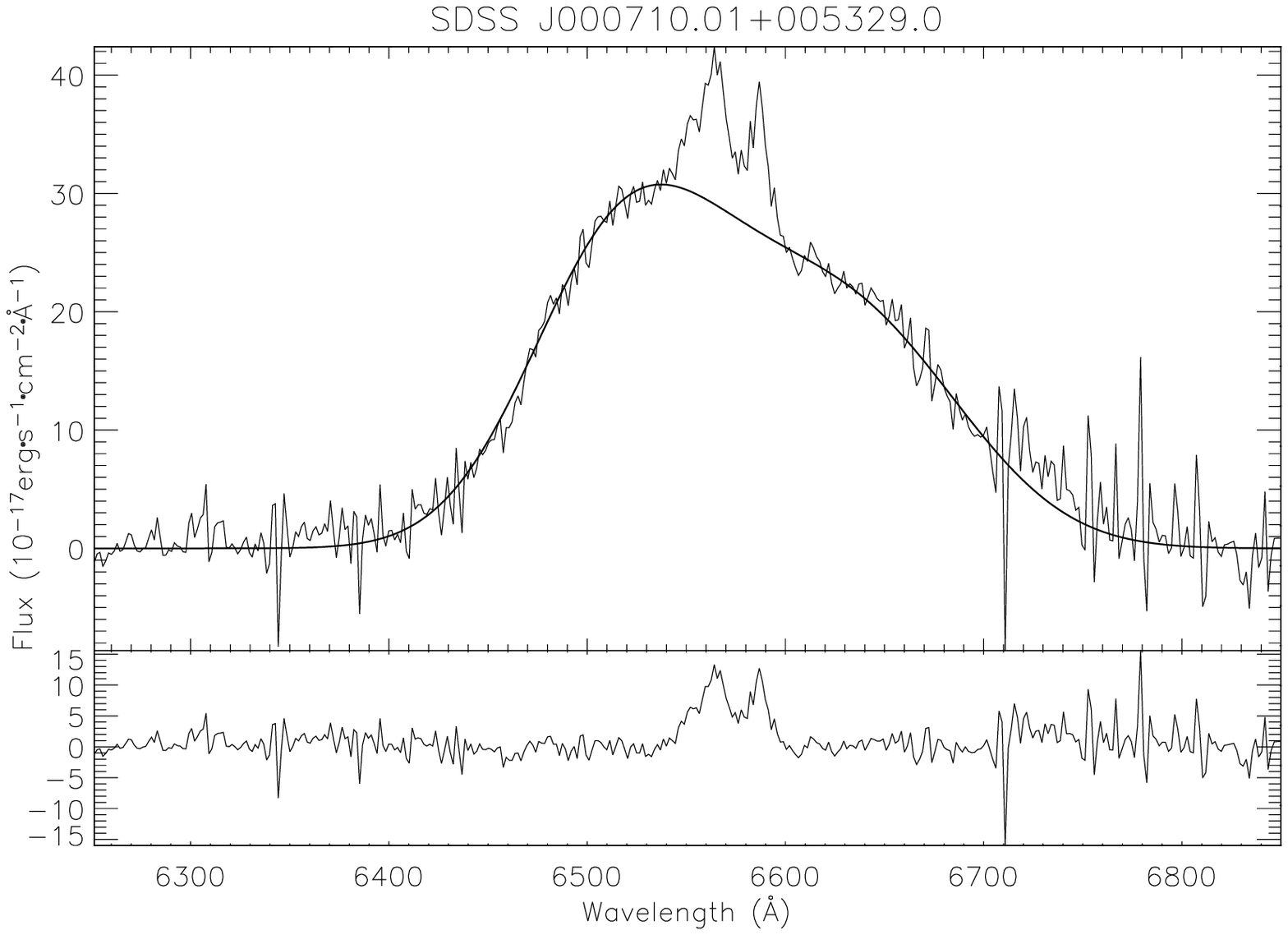}
\includegraphics[height=5cm,width=84mm]{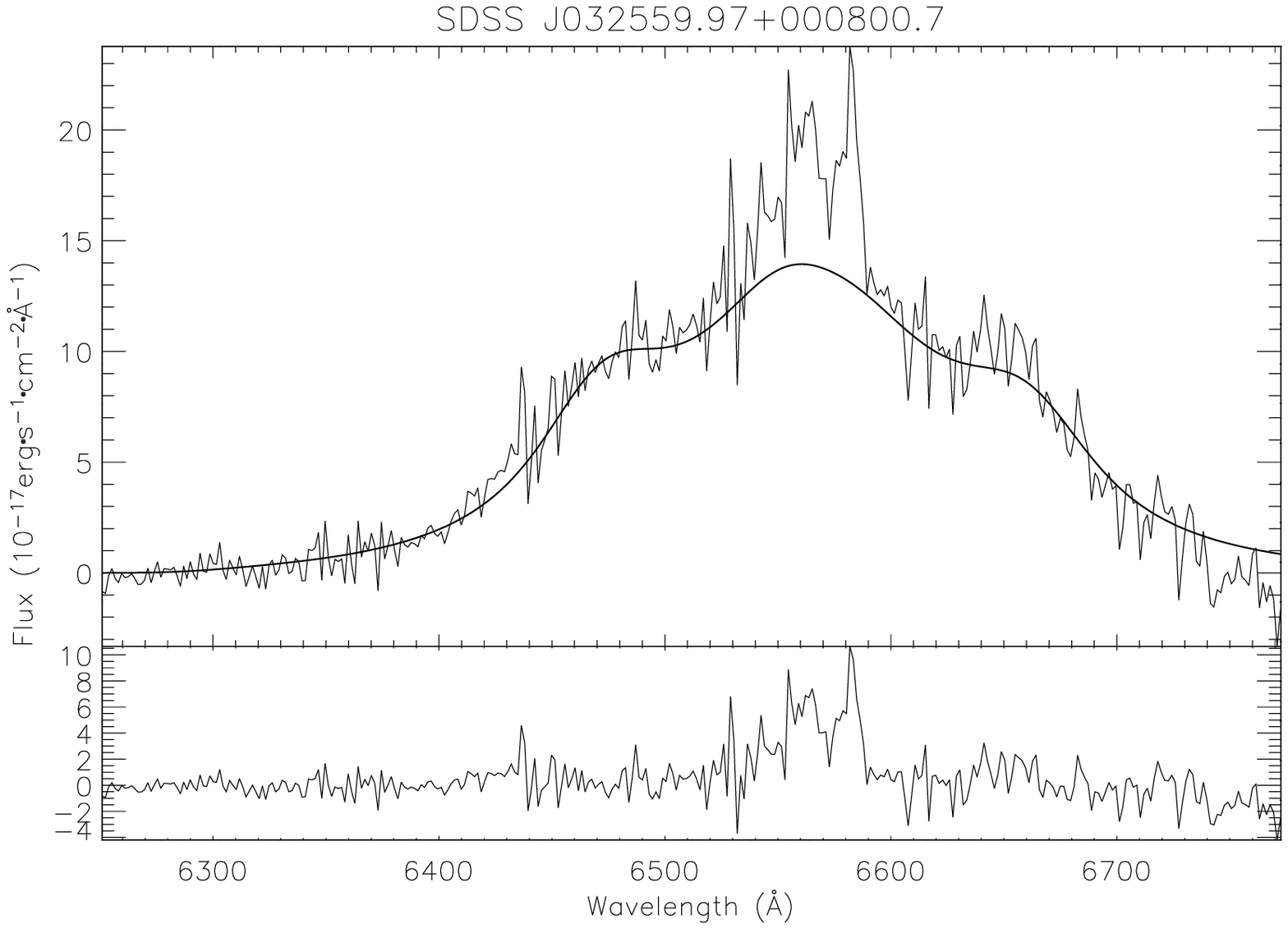}
\includegraphics[height=5cm,width=84mm]{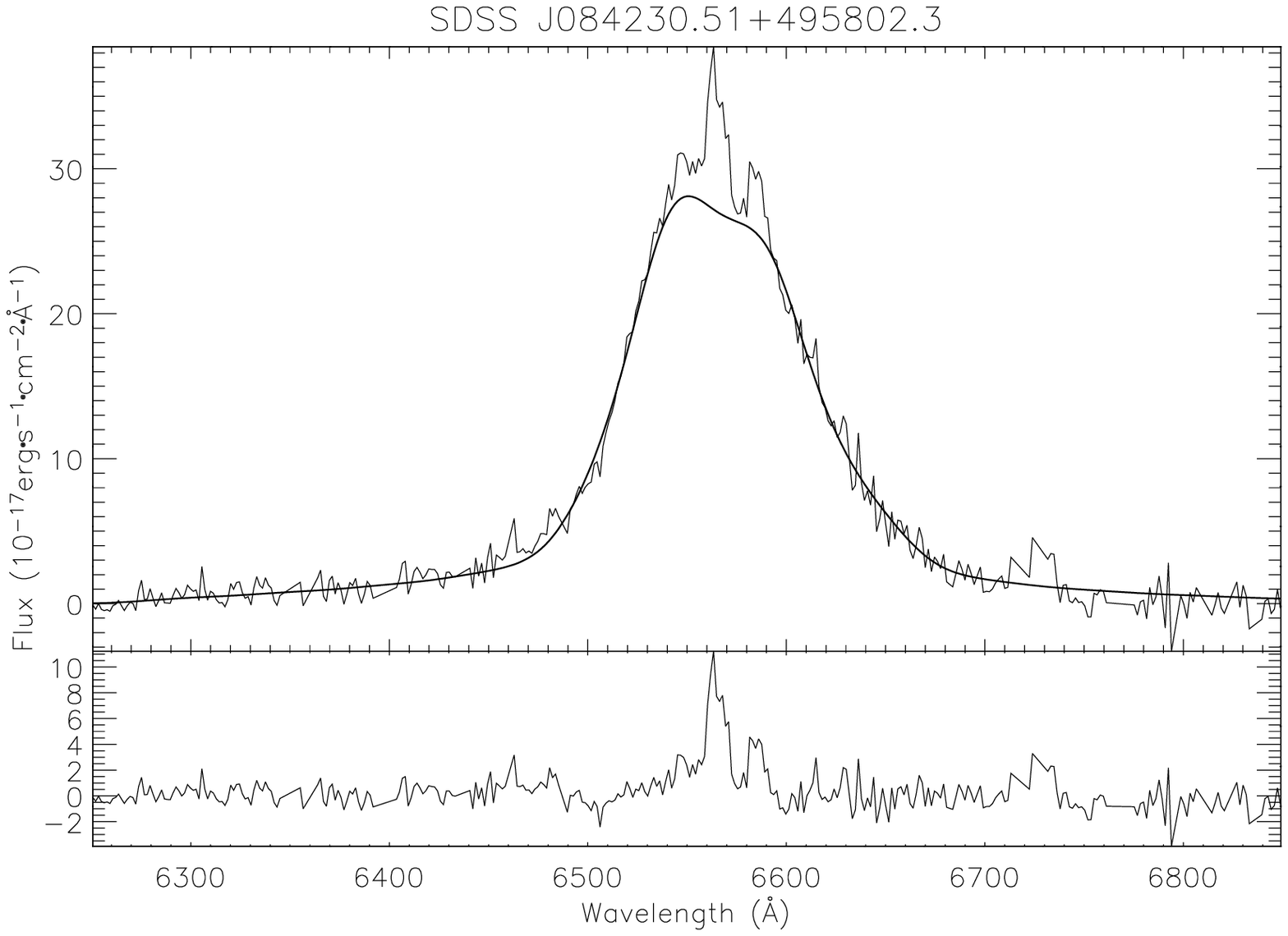}
\includegraphics[height=5cm,width=84mm]{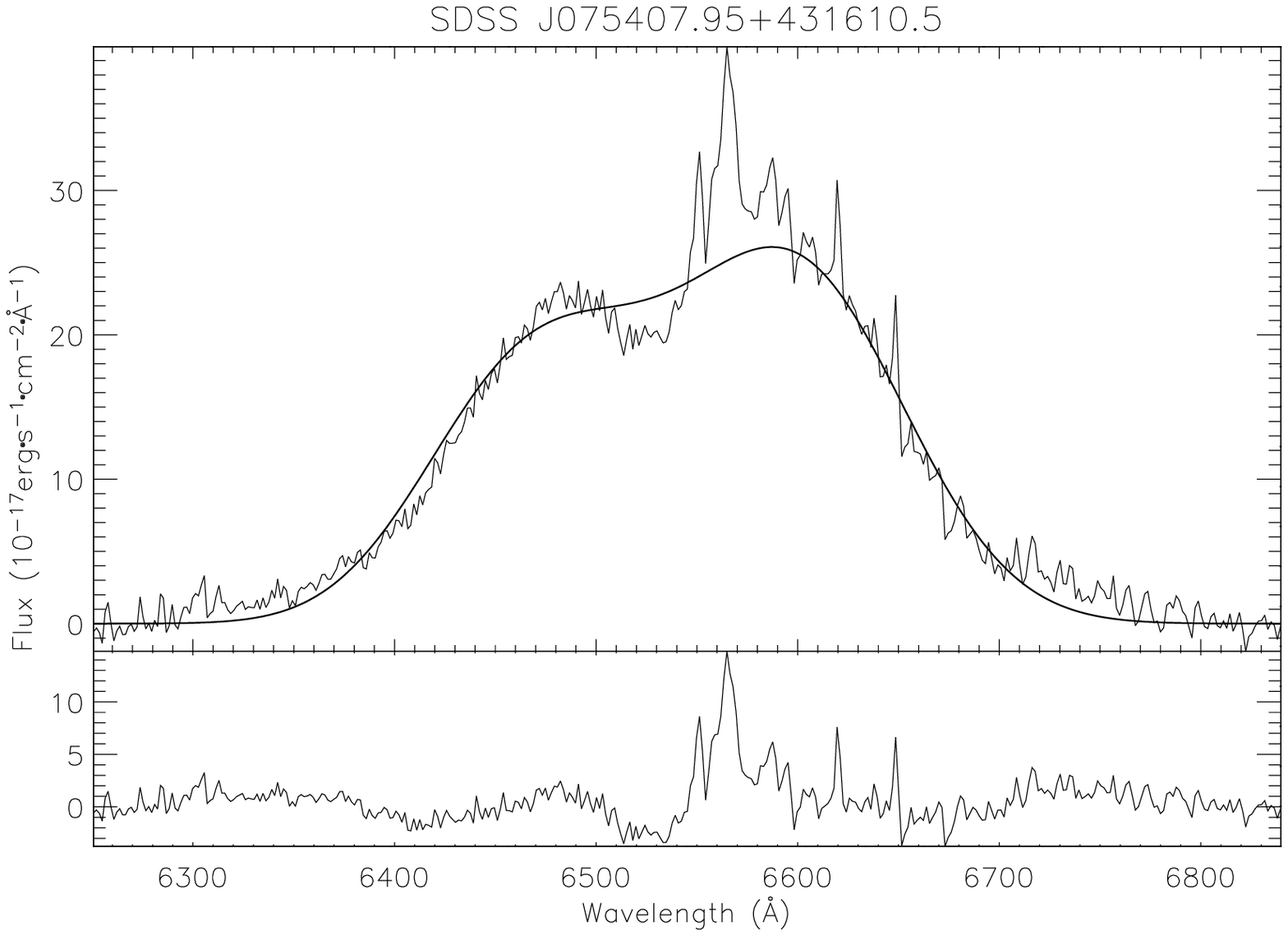}
\includegraphics[height=5cm,width=84mm]{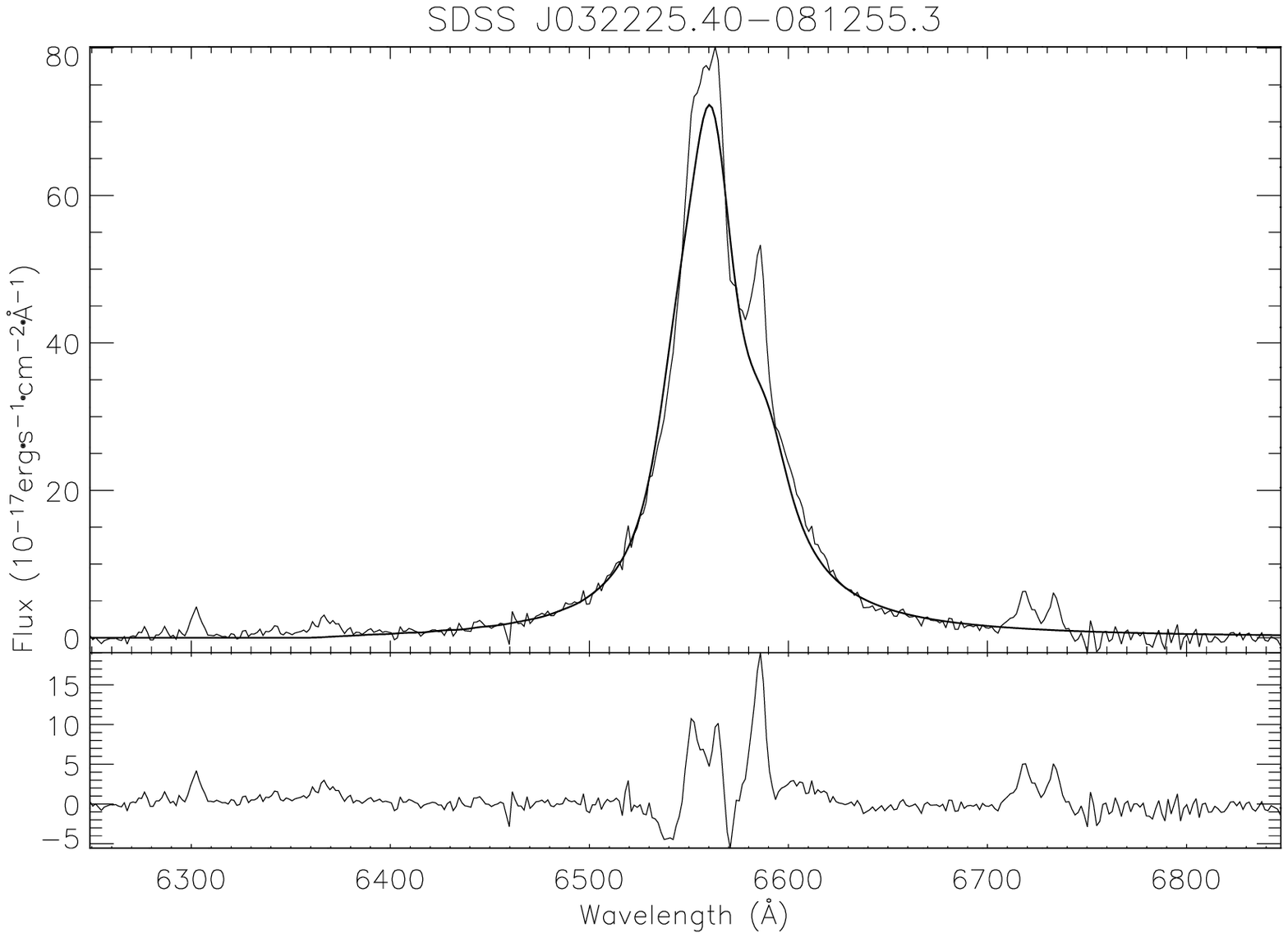}
\includegraphics[height=5cm,width=84mm]{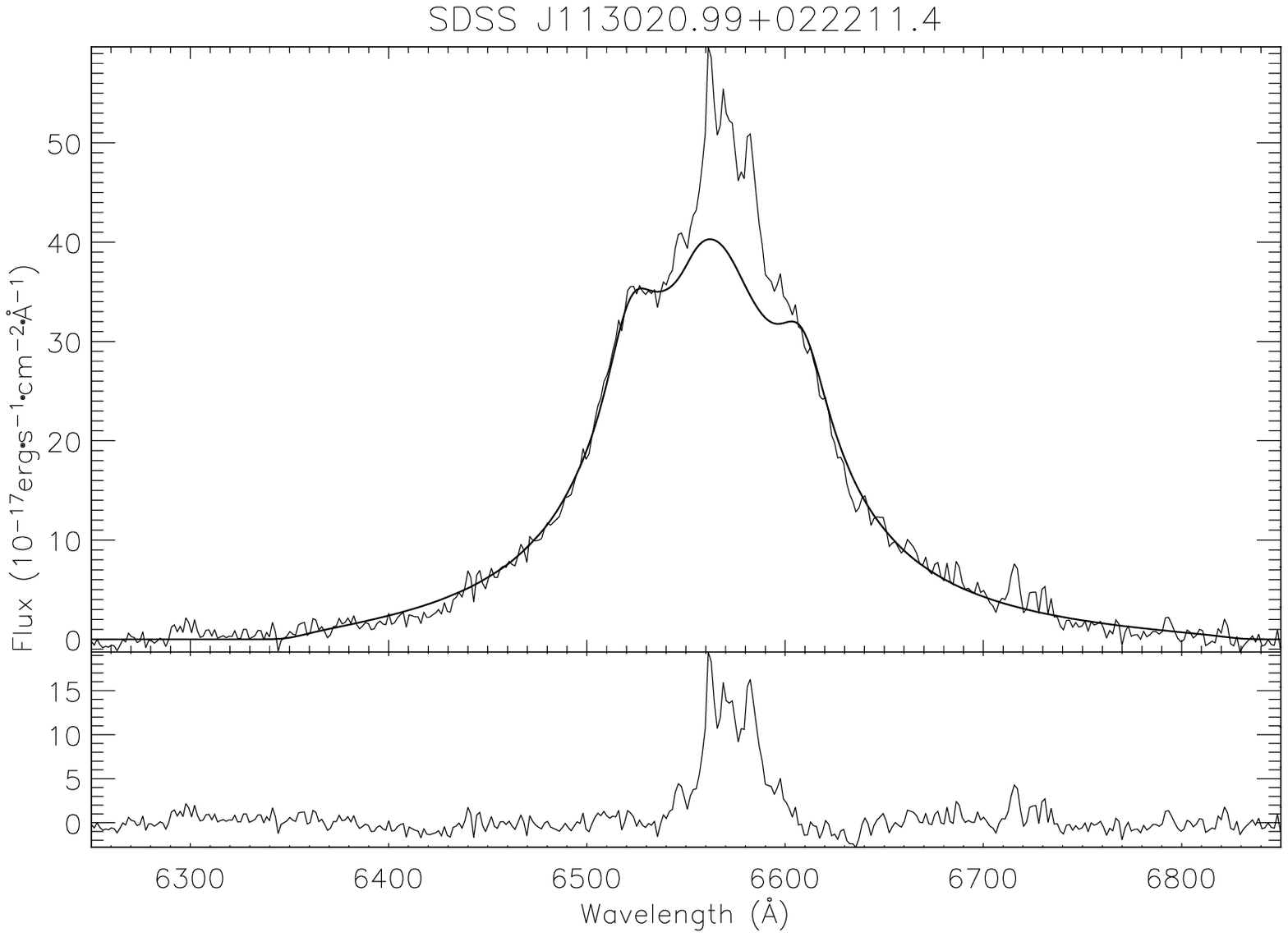}
\includegraphics[height=5cm,width=84mm]{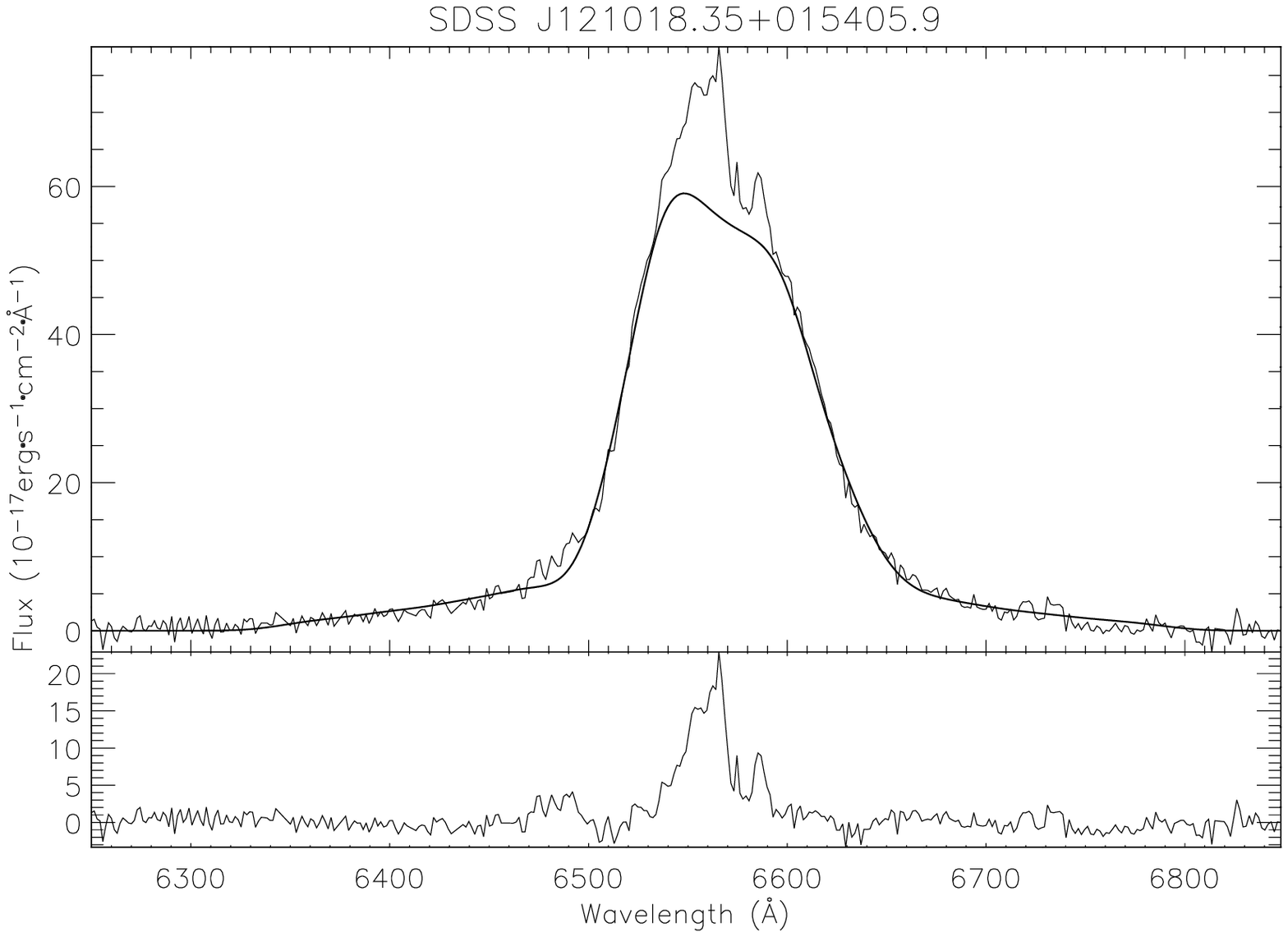}
\includegraphics[height=5cm,width=84mm]{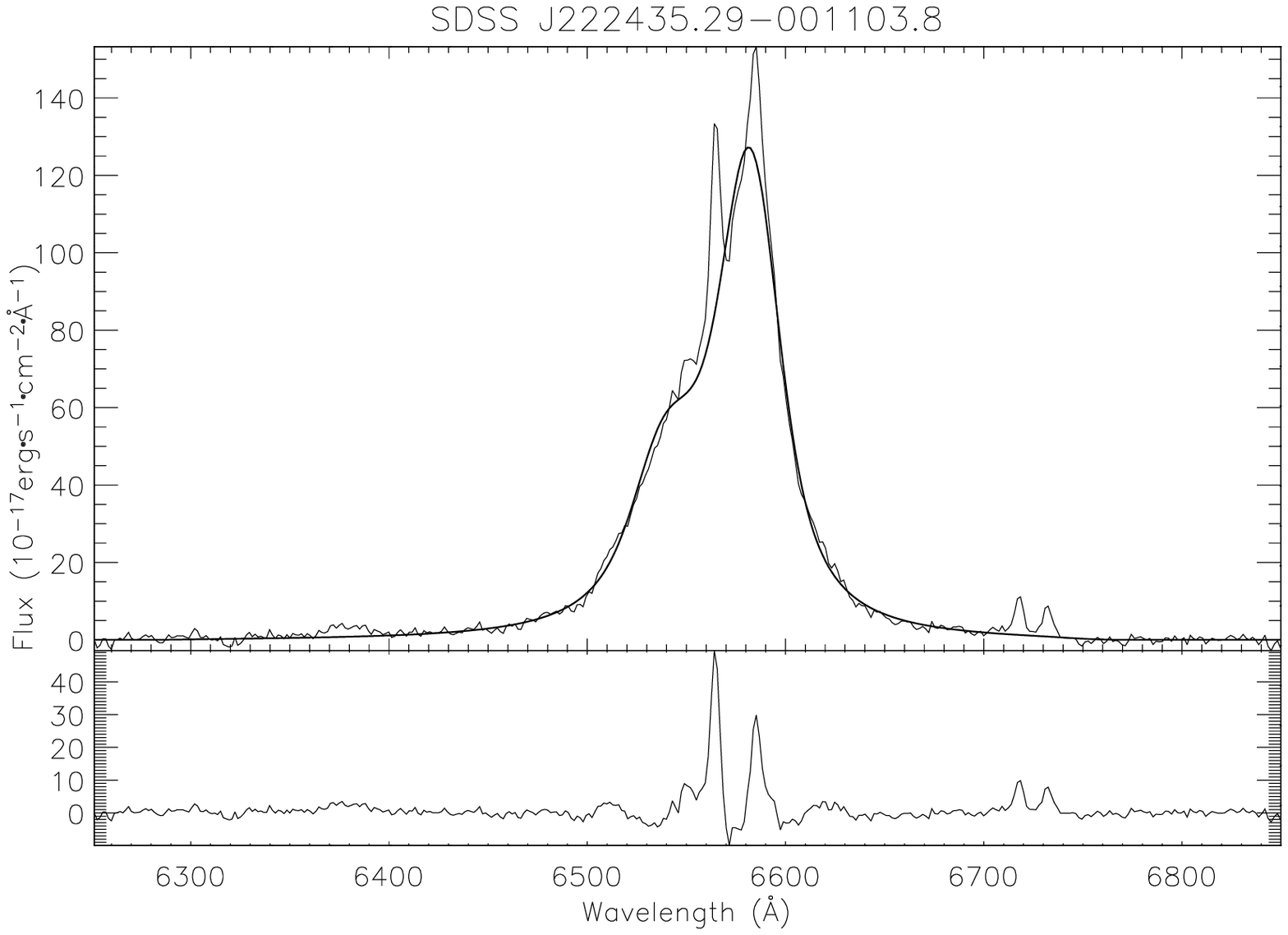}
\caption{The best fitted results for double-peaked broad H$\alpha$.
The lower panel in each plot shows the spectrum after the
subtraction of the broad H$\alpha$ coming from the accretion disk.
The model parameters are listed in Table 1. Here, only the first eight
objects are shown in the figure.}
\end{figure}

\newpage
\twocolumn
\begin{figure}
\centering\includegraphics[height = 6.6cm,width = 84mm]{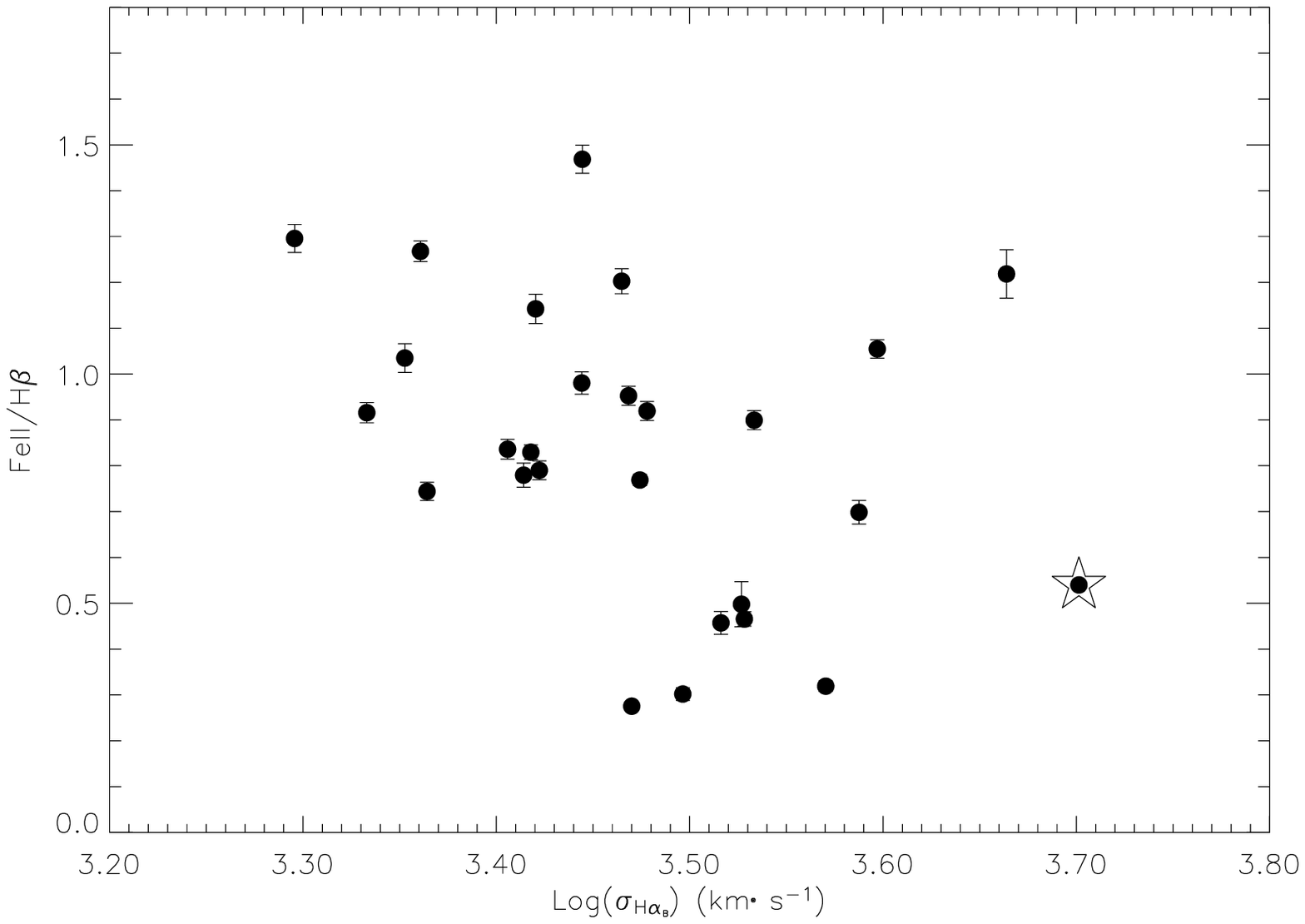}
\caption{The correlation between EW(FeII)/EW(H$\beta$) and the line width of
broad H$\alpha$. The five-point
star represents the object SDSS J2125-0813.
}
\end{figure}

\begin{figure}
\centering\includegraphics[height = 6.6cm,width = 84mm]{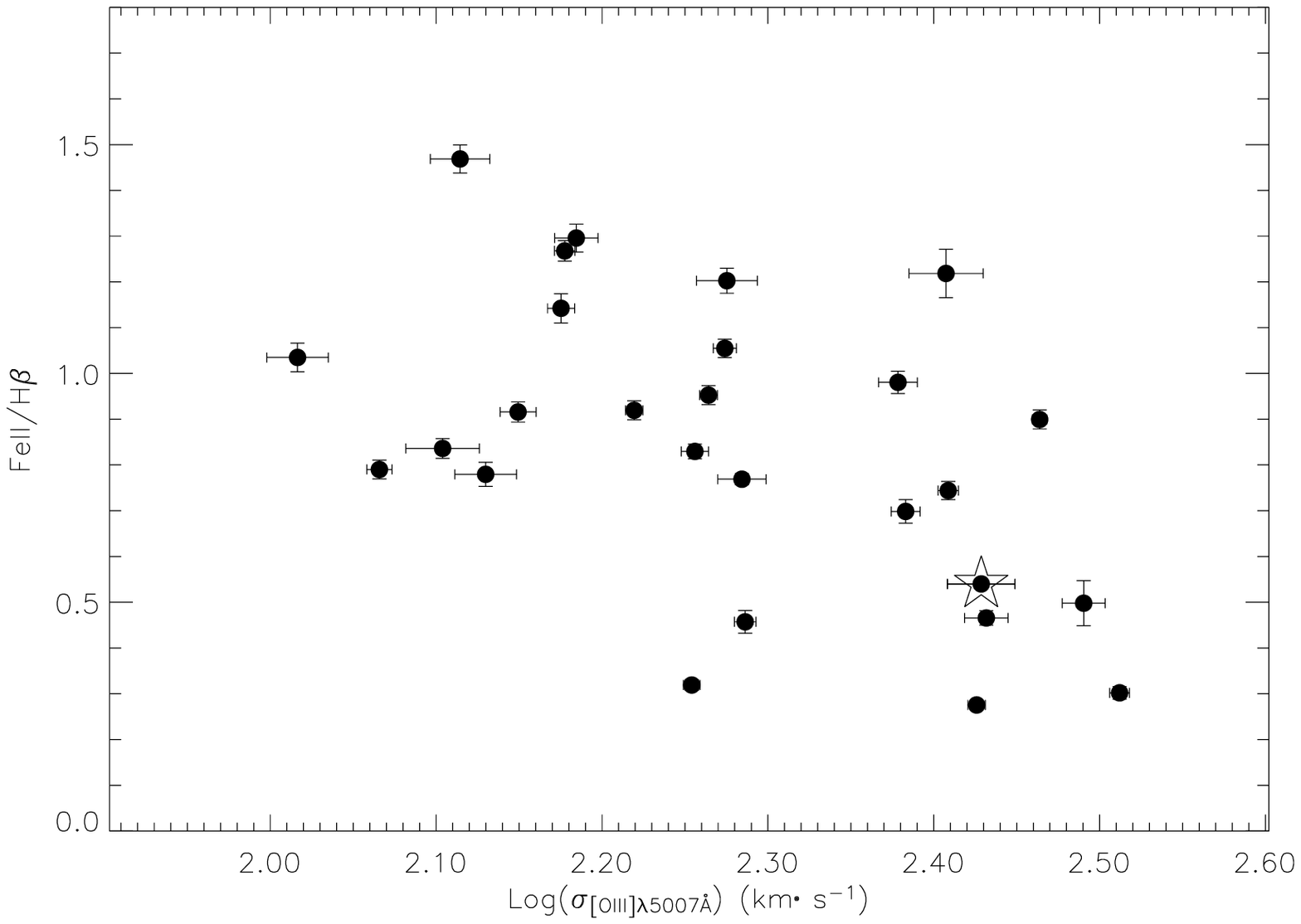}
\caption{The correlation between EW(FeII)/EW(H$\beta$) and
the line width of [OIII]$\lambda5007\AA$. The five-point
star represents the object SDSS J2125-0813.
}
\end{figure}

\begin{figure}
\centering\includegraphics[height = 6.6cm,width = 84mm]{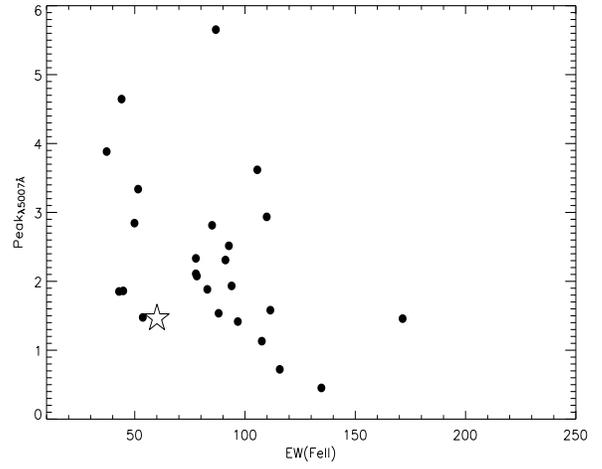}
\caption{The correlation between EW(FeII) and the height ratio of
[OIII]$\lambda5007\AA$ to that of H$\beta$. The five-point
star represents the object SDSS J2125-0813.}
\end{figure}

\begin{figure}
\centering\includegraphics[height = 6.6cm,width = 84mm]{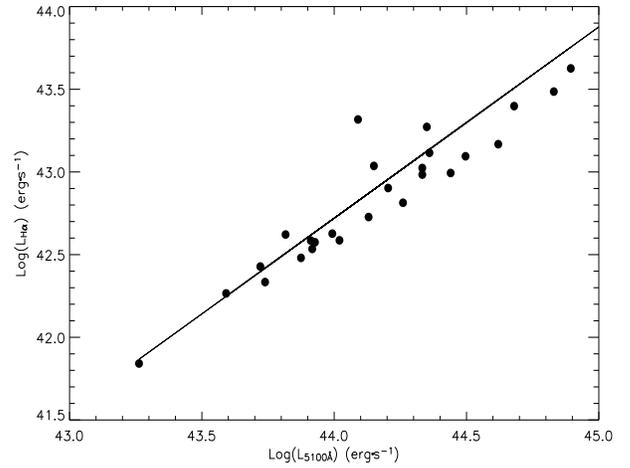}
\caption{The correlation between the continuum luminosity and the luminosity
of H$\alpha$.}
\end{figure}

\begin{figure}
\centering\includegraphics[height = 6.6cm,width = 84mm]{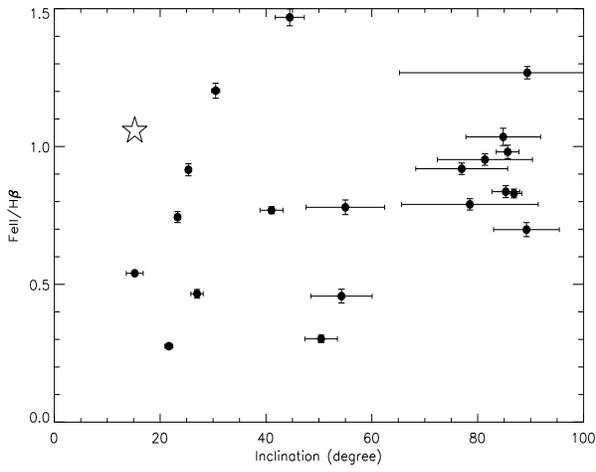}
\caption{The correlation between EW(FeII)/EW(H$\beta$) and the inclination angle
of the accretion disk. The five-point
star represents the object SDSS J2125-0813.}
\end{figure}
\begin{figure}
\centering\includegraphics[height = 6.6cm,width = 84mm]{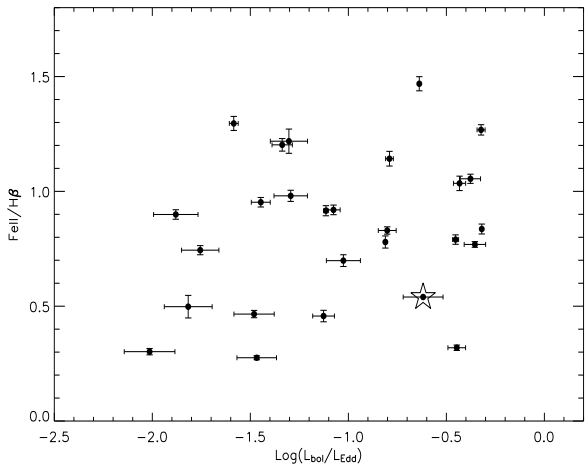}
\caption{The correlation between accretion rate and EW(FeII)/EW(H$\beta$).
The five-point star represents the object SDSS J2125-0813.}
\end{figure}

\newpage
\onecolumn
\begin{table*}
\centering
\caption{Data of Sample}
\begin{tabular}{llllllllll}
\hline
id & Name & $z$ & $m_r$ & $R$ & $\sigma_{[OIII]}$ &
$\sigma_{H\alpha_{B}}$ & $\log(L_{5100\AA})$ & $\log(M_{BH})$ & $\log(\dot{m})$ \\
 & & & & & ${\rm km\cdot s^{-1}}$ & ${\rm km\cdot s^{-1}}$ & ${\rm erg\cdot s^{-1}}$ & ${\rm M_{\odot}}$ & \\
\hline
0  &  J000710.01+005329.0  &  0.3162  &  17.13  &  0.48  &  290.96$\pm$2.925  &  3414.60  &  44.09  &  8.78  &  -1.88  \\
1  &  J032559.97+000800.7  &  0.3609  &  17.88  &    &  241.56$\pm$4.854  &  3868.92  &  44.62  &  8.45  &  -1.02  \\
2  &  J075407.95+431610.5  &  0.3476  &  17.23  &  1.50  &  179.50$\pm$2.068  &  3718.52  &  44.68  &  7.94  &  -0.44  \\
3  &  J113020.99+022211.4  &  0.2411  &  17.51  &  0.45u  &  325.12$\pm$4.467  &  3136.51  &  44.15  &  8.97  &  -2.01  \\
4  &  J001224.02-102226.2  &  0.2282  &  16.91  &  0.65  &  270.19$\pm$8.137  &  3375.07  &  44.36  &  8.65  &  -1.48  \\
5  &  J233254.46+151305.4  &  0.2147  &  17.23  &    &  193.32$\pm$2.888  &  3282.24  &  44.13  &  8.07  &  -1.12  \\
6  &  J091828.59+513932.1  &  0.1854  &  17.13  &  0.43  &  149.72$\pm$2.815  &  2632.22  &  44.02  &  7.62  &  -0.79  \\
7  &  J093653.84+533126.9  &  0.2280  &  16.86  &  0.53  &  266.58$\pm$3.254  &  2951.40  &  44.35  &  8.63  &  -1.46  \\
8  &  J133957.99+613933.4  &  0.3724  &  18.68  &    &  309.30$\pm$9.214  &  3363.38  &  44.26  &  8.89  &  -1.81  \\
9  &  J140019.27+631426.9  &  0.3314  &  17.76  &  0.55u  &  255.49$\pm$13.13  &  4611.71  &  44.44  &  8.55  &  -1.30  \\
10  &  J154019.57-020505.4  &  0.3204  &  16.73  &  0.80  &  187.93$\pm$3.000  &  3953.81  &  44.83  &  8.02  &  -0.37  \\
11  &  J2125-0813           &  0.6239  &  17.07  &  0.46  &  268.29$\pm$12.57  &  5026.95  &  45.21  &  8.64  &  -0.61  \\
12  &  J084230.51+495802.3  &  0.3050  &  17.93  &  0.59u  &  239.02$\pm$6.430  &  2780.96  &  44.33  &  8.44  &  -1.29  \\
13  &  J032225.40-081255.3  &  0.1260  &  17.96  &    &  141.04$\pm$3.522  &  2152.76  &  43.59  &  7.52  &  -1.11  \\
14  &  J121018.35+015405.9  &  0.2158  &  17.23  &  0.30u  &  180.30$\pm$3.434  &  2617.29  &  44.33  &  7.94  &  -0.80  \\
15  &  J222435.29-001103.8  &  0.0581  &  17.05  &    &  152.92$\pm$4.589  &  1975.48  &  43.26  &  7.66  &  -1.58  \\
16  &  J103421.70+605318.1  &  0.2277  &  17.65  &  0.60u  &  127.02$\pm$6.491  &  2545.75  &  44.20  &  7.33  &  -0.31  \\
17  &  J222055.73-075317.9  &  0.1489  &  17.60  &    &  188.47$\pm$7.977  &  2916.25  &  43.87  &  8.02  &  -1.33  \\
18  &  J102914.87+572353.7  &  0.1885  &  17.79  &  0.53u  &  256.27$\pm$3.611  &  2312.89  &  43.99  &  8.56  &  -1.75  \\
19  &  J094859.47+433518.9  &  0.2262  &  16.91  &  0.16u  &  150.49$\pm$2.132  &  2294.98  &  44.49  &  7.63  &  -0.32  \\
20  &  J100756.46+491809.5  &  0.1496  &  17.98  &  0.60u  &  103.85$\pm$4.431  &  2252.58  &  43.73  &  6.98  &  -0.43  \\
21  &  J095744.43+075124.8  &  0.1406  &  17.61  &  0.47u  &  134.84$\pm$5.772  &  2594.93  &  43.81  &  7.44  &  -0.81  \\
22  &  J140502.65+470747.5  &  0.1521  &  17.54  &  0.42u  &  130.14$\pm$5.365  &  2782.77  &  43.92  &  7.37  &  -0.63  \\
23  &  J151535.25+480530.5  &  0.3115  &  16.52  &  0.01u  &  192.45$\pm$6.467  &  2980.23  &  44.89  &  8.06  &  -0.35  \\
24  &  J144207.69+384411.3  &  0.1457  &  17.48  &  0.40u  &  116.35$\pm$2.027  &  2643.79  &  43.91  &  7.18  &  -0.45  \\
25  &  J153415.41+303435.4  &  0.0938  &  16.98  &  0.20u  &  183.73$\pm$2.272  &  2940.21  &  43.72  &  7.98  &  -1.44  \\
26  &  J082930.59+081238.0  &  0.1291  &  17.21  &  0.32  &  165.72$\pm$1.981  &  3005.78  &  43.91  &  7.80  &  -1.07  \\
\hline
\end{tabular}\\
Notes:\\
The first Column lists the ID number of the object.
The second and third Columns present the name in the format of "SDSS
Jhhmmss.ss$\pm$ddmmss.s" and redshift of each object.
The forth column presents the PSF magnitude in $r$ band of SDSS.
The fifth column gives the value of $R_r$ (radio loudness) as defined by
Ivezi$\rm{\acute{c}}$ et al. (2002), 'u' represents the value of an
upper limit one.
The following two columns list the line width of
[NII]$\lambda6583\AA$ and broad H$\alpha$ in units of ${\rm km\cdot s^{-1}}$.
The last three columns contain the value of continuum
luminosity at 5100$\AA$ (corrected for Galactic extinction including
the redden correction, following Schlegel et al., 1998), the value
of BH masses estimated from the line width of [OIII]$\lambda5007\AA$
(as a tracer of velocity dispersion) and the dimensionless accretion
rate of the dbp emitters.\\
The line width of broad H$\alpha$ of J2125-0813 is estimated from
the linewidth of broad H$\beta$ produced by the
accretion disk model. Here, we just list the PSF magnitude at r band
for each object.
\end{table*}

\begin{table*}
\centering
\caption{Parameters of the accretion disk model}
\begin{tabular}{lllllllll}
\hline
id & $r_{in}$ & $r_{out}$ &  $i$ & $e$ &
$q$ & $\sigma$
& $k_{H\beta}$ & $k_{FeII}$ \\
 & ${\rm R_{g}}$ & ${\rm R_{g}}$ &  & &
 & ${\rm km\cdot s^{-1}}$
&  &  \\
\hline
0  &  991  &  4054  &  33  &  0.25  &  2.57  &  2032$\pm$754  &  0.39  &  0.35  \\
1  &  391$\pm$62  &  4085$\pm$151  &  89$\pm$6  &  0.26$\pm$0.01  &  0.98$\pm$0.11  &  738$\pm$90  &  0.40  &  0.28  \\
2  &  385  &  969  &  28  &  0.57  &  3.61  &  2059$\pm$400  &  0.43  &  0.13  \\
3  &  367$\pm$37  &  11946$\pm$1077  &  50$\pm$3  &  0.24$\pm$0.01  &  1.71$\pm$0.01  &  296$\pm$25  &  0.37  &  0.11  \\
4  &  37$\pm$17  &  2660$\pm$158  &  26$\pm$1  &  0.21$\pm$0.01  &  1.29$\pm$0.02  &  700$\pm$43  &  0.39  &  0.18  \\
5  &  526$\pm$89  &  5557$\pm$833  &  54$\pm$5  &  0.21$\pm$0.01  &  1.29$\pm$0.08  &  771$\pm$60  &  0.47  &  0.21  \\
6  &  526$\pm$21  &  3350$\pm$255  &  89  &  0.96$\pm$0.01  &  2.24$\pm$0.05  &  703$\pm$48  &  0.36  &  0.41  \\
7  &  103$\pm$3  &  1804$\pm$29  &  21$\pm$0  &  0.27$\pm$0.02  &  1.21$\pm$0.01  &  915$\pm$21  &  0.36  &  0.10  \\
8  &  2797  &  >32768  &  89  &  0.17  &  7.40$\pm$1.86  &  1622$\pm$275  &  0.54  &  0.27  \\
9  &  210$\pm$112  &  5705  &  47  &  0.03  &  1.43$\pm$0.20  &  1813$\pm$1078  &  0.34  &  0.42  \\
10  &  394$\pm$40  &  4953$\pm$142  &  89  &  0.24$\pm$0.01  &  0.97$\pm$0.07  &  1238$\pm$55  &  0.34  &  0.35  \\
11  &  31$\pm$6  &  339$\pm$40  &  15$\pm$1  &  0.01  &  1.58$\pm$0.19  &  2610$\pm$270  &  1.00  &  0.53  \\
12  &  222$\pm$18  &  3382$\pm$374  &  85$\pm$2  &  0.90$\pm$0.01  &  2.03$\pm$0.04  &  624.6$\pm$46.5  &  0.45  &  0.44  \\
13  &  30$\pm$8  &  12790$\pm$507  &  25$\pm$0  &  0.33$\pm$0.01  &  1.56$\pm$0.01  &  292$\pm$13  &  0.42  &  0.38  \\
14  &  408$\pm$11  &  3100$\pm$116  &  86$\pm$1  &  0.94$\pm$0.01  &  2.15$\pm$0.02  &  562.5$\pm$40.5  &  0.51  &  0.42  \\
15  &  578$\pm$340  &  >32768  &  72  &  0.26$\pm$0.08  &  1.36$\pm$0.01  &  479$\pm$16  &  0.36  &  0.47  \\
16  &  208$\pm$12  &  4830$\pm$333  &  85$\pm$2  &  0.84$\pm$0.01  &  1.85$\pm$0.02  &  499$\pm$24  &  0.44  &  0.36  \\
17  &  11$\pm$1  &  7779$\pm$428  &  30$\pm$0  &  0.23$\pm$0.01  &  1.44$\pm$0.01  &  476.4$\pm$33.30  &  0.54  &  0.65  \\
18  &  23$\pm$13  &  6843$\pm$387  &  23$\pm$0  &  0.34$\pm$0.01  &  1.54$\pm$0.03  &  461.4$\pm$27.00  &  0.53  &  0.40  \\
19  &  88$\pm$7  &  3401$\pm$226  &  89$\pm$2  &  0.84$\pm$0.01  &  1.46$\pm$0.02  &  572.7$\pm$21.30  &  0.41  &  0.52  \\
20  &  406$\pm$30  &  4875$\pm$346  &  84$\pm$7  &  0.87$\pm$0.01  &  1.92$\pm$0.04  &  451.5$\pm$32.40  &  0.51  &  0.53  \\
21  &  141$\pm$43  &  26691$\pm$4774  &  54$\pm$7  &  0.28$\pm$0.01  &  1.50$\pm$0.01  &  429.6$\pm$23.70  &  0.30  &  0.23  \\
22  &  99$\pm$19  &  15609$\pm$1578  &  44$\pm$2  &  0.32$\pm$0.01  &  1.63$\pm$0.01  &  303.9$\pm$23.70  &  0.51  &  0.75  \\
23  &  104$\pm$16  &  13542$\pm$1252  &  41$\pm$2  &  0.33$\pm$0.01  &  1.73$\pm$0.01  &  461.7$\pm$27.30  &  0.54  &  0.42  \\
24  &  99$\pm$13  &  2551$\pm$278  &  78$\pm$1  &  0.81$\pm$0.01  &  1.54$\pm$0.03  &  566.1$\pm$22.80  &  0.50  &  0.40  \\
25  &  307$\pm$24  &  21499$\pm$1118  &  81$\pm$8  &  0.28$\pm$0.01  &  1.56$\pm$0.01  &  352.2$\pm$22.50  &  0.44  &  0.42  \\
26  &  77$\pm$6  &  2285$\pm$150  &  76$\pm$8  &  0.83$\pm$0.01  &  1.71$\pm$0.02  &  676.8$\pm$29.10  &  0.39  &  0.36  \\
\hline
\end{tabular}\\
Notes:\\
If the error of the parameter is larger than the value of the parameter,
we do not show the error. The id number represents the name in the Table 1.
\end{table*}
\end{document}